%% file: main.tex
\documentclass[final,3p,times]{elsarticle}
\usepackage[ruled,vlined]{algorithm2e}
\usepackage{amsfonts}
\usepackage{amsmath}
\usepackage{amssymb}
\usepackage{mathtools}
\usepackage{booktabs}
\usepackage{arydshln}
\usepackage{graphicx}
\usepackage{multirow}
\usepackage{xcolor}
\usepackage[colorlinks=true,citecolor=blue]{hyperref}
\usepackage{chngcntr}
\usepackage{float}
\usepackage{bm}
\usepackage{bbm}
\newcommand\myvec[1]{{\bm{#1}}}

\journal{}
\begin{document}

\begin{frontmatter}

\title{Nested Fourier-enhanced neural operator for efficient modeling of radiation transfer in fires}

\author[1]{Anran Jiao}
\author[2]{Wengyao Jiang}
\author[3,*]{Xiaoyi Lu}
\author[3]{Yi Wang}
\author[1,4,**]{Lu Lu}

\cortext[*]{Corresponding author. \textit{E-mail address}: xiaoyi.lu@fm.com}
\cortext[**]{Corresponding author. \textit{E-mail address}:  lu.lu@yale.edu}

\address[1]{Department of Statistics and Data Science, Yale University, New Haven, CT 06511, USA}
\address[2]{Department of Biostatistics, Yale University, New Haven, CT 06511, USA}
\address[3]{FM, Research Division, Norwood, MA 02062, USA}
\address[4]{Department of Chemical and Environmental Engineering, Yale University, New Haven, CT 06511, USA}

%====================================================================
\begin{abstract}

Computational fluid dynamics (CFD) has become an essential tool for predicting fire behavior, yet maintaining both efficiency and accuracy remains challenging. A major source of computational cost in fire simulations is the modeling of radiation transfer, which is usually the dominant heat transfer mechanism in fires. Solving the high-dimensional governing radiative transfer equation (RTE) with traditional numerical methods can be a performance bottleneck. Here, we present a machine learning framework based on Fourier-enhanced multiple-input neural operators (Fourier-MIONet) as an efficient alternative to direct numerical integration of the RTE. We first investigate the performance of neural operator architectures for a small-scale 2D pool fire and find that Fourier-MIONet provides the most accurate radiative solution predictions. The approach is then extended to 3D CFD fire simulations, where the computational mesh is locally refined across multiple levels. In these high-resolution settings, monolithic surrogate models for direct field-to-field mapping become difficult to train and computationally inefficient. This lack of scalability makes the use of such neural surrogates impractical for engineering CFD applications. To address this issue, a nested Fourier-MIONet is proposed to train and infer radiation solutions across multiple mesh-refinement levels. Validation of the approach is conducted on 3D McCaffrey pool fires simulated using the CFD solver, FireFOAM, including fixed fire sizes (characterized by heat release rate, HRR) and a single unified model trained to generalize across a continuous range of fire sizes. The proposed method achieves global relative errors of 2--4\% for 3D varying-HRR scenarios while providing faster inference than the estimated cost of one finite-volume radiation solve in FireFOAM for the 16-solid-angle case considered here. With fast and accurate inference, the surrogate makes higher-fidelity radiation treatments practical and enables the incorporation of more spectrally resolved radiation models into CFD fire simulations for engineering applications.

\end{abstract}

\end{frontmatter}

%====================================================================

\section{Introduction}
\label{sec:intro}

Fire is a leading driver of property loss and business interruption across industrial and commercial occupancies. Fire risk resilience requires both an understanding of fire dynamics and predictive tools to evaluate potential fire scenarios. Computational fire dynamics modeling, typically based on large-eddy simulation (LES) coupled with submodels for combustion, turbulence, and heat transfer, is increasingly used to support this effort and complement physical fire tests~\cite{Wang11PROCI, McGrattan10FDS}. In fire scenarios, thermal radiation often dominates heat transfer from flames and hot, sooty plumes to the surroundings~\cite{agarwal2021fire}. Modeling radiative heat transfer is therefore essential for reliable hazard assessment and for applying computational fluid dynamics (CFD) to develop and evaluate fire protection strategies.

The radiative transfer equation (RTE) governs the propagation of thermal radiation in participating media, accounting for emission, absorption, and scattering. Traditional numerical methods for solving the RTE include the finite volume method (FVM)~\cite{raithby1990finite} and the discrete ordinates method (DOM)~\cite{thynell1998discrete}, which are widely used in fire CFD codes like FireFOAM~\cite{Wang11PROCI} and Fire Dynamics Simulator (FDS)~\cite{McGrattan10FDS}. These methods solve the integro-differential equation by discretizing the angular domain into a finite set of solid angles. While readily coupled with CFD solvers, these methods are prone to ray effects unless a high-resolution angular discretization is employed, which imposes high computational and memory costs. Alternatively, the P$_n$ (e.g., P1 model) approximation reduces the RTE to a set of simplified elliptic partial differential equations by expanding the intensity into lowest-order spherical harmonics. Although computationally efficient, the P1 model assumes a nearly isotropic radiation field, leading to substantial inaccuracies in optically thin media and regions with severe directional gradients. Monte Carlo ray tracing (MCRT)~\cite{mahan2018monte} avoids angular discretization errors but suffers from slow statistical convergence, making it impractical for large-scale transient simulations. Thus, there remains a need for efficient and accurate RTE solution procedures.

Recent advances in neural network-based surrogate modeling have expanded viable options for computational radiation modeling. Physics-informed neural networks (PINNs) provide a framework for solving partial differential equations (PDEs) by embedding the residual directly into the loss function~\cite{dissanayake1994neural, lagaris1998artificial, raissi2019pinn,pang2019fpinns, karniadakis2021physics, lu2021deepxde, lu2021physics,yu2022gradient,wu2023comprehensive}, and have been successfully applied to solving forward as well as inverse problems of PDEs across various fields \cite{chen2020physics,yazdani2020systems,daneker2023systems,fan2025deep,daneker2024transfer}. Mishra and Molinaro~\cite{Mishra21JQSRT} used this methodology to solve the RTE by employing a neural network that takes spatial coordinates, time, angular directions, and frequencies as inputs to predict the radiative intensity. They demonstrated that PINNs can approximate the intensity field and handle inverse problems, such as identifying unknown absorption coefficients from incident radiation measurements. While this approach shows the potential of PINNs for radiative transfer, 
these models typically solve for a single PDE instance for a specific set of configuration parameters. In transient CFD simulations, the parameters and input functions of governing PDEs evolve at every time step. Therefore, a neural surrogate solver must generalize across varying input fields, which PINNs fall short of without retraining. 

Deep neural operators such as deep operator networks (DeepONets)~\cite{deeponetNatureML, lu2022comprehensive,lu2022multifidelity, zhu2023reliable} address this limitation by learning nonlinear operator mappings between function spaces. By incorporating physics-informed losses~\cite{wang2021learning, jiao2024solving}, the physics-informed DeepONet (PI-DeepONet) can be developed to learn solution operators across varying input parameters. Lu and Wang~\cite{Lu24PROCI} applied a physics-informed version of the multiple-input operator network (MIONet)~\cite{jin2022mionet} to the RTE by treating absorption and emission fields as separate input functions. This framework successfully predicts radiative source terms in non-gray gas problems and has been extended to complex pool fire scenarios~\cite{Sorokin24NIPS}. However, a significant challenge arises when scaling these operators to 3D engineering applications where radiation is resolved on high-resolution, locally refined meshes.

In this study, we pursue a data-driven framework to resolve these scaling issues and develop models that are both accurate and computationally efficient for practical 2D and 3D pool fire simulations. The input fields defined on CFD meshes comprise hundreds of thousands of nodal or cell values in our 3D cases, creating a high-dimensional input space that is difficult to process. We address this by employing principal component analysis (PCA) to project high-dimensional fields onto their principal components, effectively reducing the parameter space while preserving the critical physical features of the plume. Also, fires are often turbulent and exhibit multi-scale features, making it difficult for conventional multi-layer perceptrons (MLPs) to learn fine-scale features as they tend to capture low-frequency, smooth components more easily than high-frequency details. We thus explore the integration of Kolmogorov-Arnold networks (KAN)~\cite{liu2024kan} and Fourier-enhanced layers~\cite{zhu2023fourier, jiang2024fourier}. These architectures, which we refer to as PCA-MIONet-KAN and Fourier-MIONet, improve approximation capabilities for radiation in turbulent fires. Furthermore, practical fire simulations often utilize meshes with multiple levels of local refinement to resolve turbulent fire dynamics. This results in a complex data structure where the mesh is locally structured within each refinement level but globally non-aligned. A single Fourier-MIONet cannot efficiently process this heterogeneity in the mesh. The nested Fourier-DeepONet~\cite{wen2023real,lee2024efficient} has been proposed to produce predictions at different refinement levels using a hierarchy of DeepONets for a geological carbon sequestration application. We adopt a similar nested Fourier-MIONet architecture in this study.

We present a Fourier-MIONet framework for learning the solution operators of the RTE in 2D and 3D fire scenarios. The objective is to provide an efficient and accurate surrogate for radiative transfer across different fires. We develop models for both constant heat release rate (HRR) cases and a single unified model that generalizes across a range of fire sizes by treating HRR as a variable parameter. The main contributions of this study are as follows.
\begin{enumerate}
\item We apply a data-driven Fourier-MIONet surrogate model that learns the solution operator of the RTE, mapping the absorption coefficient and temperature fields to the corresponding radiative intensity field. 

\item We extend this framework to a complex 3D McCaffrey fire scenario using a nested Fourier-MIONet that operates across multiple mesh-refinement levels on constant-HRR datasets. 

\item We further demonstrate the broader applicability of our method by training a single unified model that handles variable HRR fires. To support different accuracy and speed requirements, we consider four models with different sizes that offer flexible trade-offs between inference time and predictive accuracy.
\end{enumerate}

The paper is organized as follows. In Section~\ref{sec:comp_radiation_modeling}, we introduce the computational radiation modeling and the goal of developing a neural operator surrogate for radiation. In Section~\ref{sec:methods}, we state the problem setup of RTE, present the MIONet, Fourier-MIONet and the nested Fourier-MIONet methods, and introduce the data generation. In Section~\ref{sec:results}, we demonstrate the accuracy and efficiency of Fourier-MIONet on a 2D pool fire and a 3D McCaffrey fire case.

\section{Computational radiation modeling}
\label{sec:comp_radiation_modeling}

In the computational fluid dynamics framework, radiation is coupled to the energy conservation equation. Computationally, this coupling appears as a volumetric source (or sink) term, representing the divergence of the total radiative heat flux vector, $\nabla \cdot \myvec{q}_r$. The energy equation for a reacting flow (neglecting viscous heating and external work) can be expressed in terms of sensible enthalpy $h_s$ as
\begin{equation*}
    \frac{\partial (\rho h_s)}{\partial t} + \nabla \cdot (\rho \myvec{u} h_s) = \nabla \cdot (\alpha \nabla h_s) + \dot{\omega}_T - \nabla \cdot \myvec{q}_r,
\end{equation*}
where $\rho$ is the density, $\myvec{u}$ is the velocity vector, $\alpha$ is the thermal diffusivity, and $\dot{\omega}_T$ represents the heat release rate from combustion. This formulation is widely adopted and implemented in open-source CFD fire simulation tools, including FireFOAM (based on OpenFOAM) \cite{Wang11PROCI} and the FDS~\cite{McGrattan10FDS}. 

The radiative source term or the divergence of the radiative flux $\myvec{q}_r(\myvec{r})$ is the spectral integral of the difference between emission and absorption
\begin{equation}
    -\nabla \cdot \myvec{q}_r = \int_0^{\infty} \kappa_\eta(\myvec{r}) \left[ G_\eta(\myvec{r}) - 4\pi I_{b,\eta}(\myvec{r}) \right] d\eta, \quad \mathrm{and}\quad G_\eta(\myvec{r}) = \int_{4\pi} I_\eta(\myvec{r}, \myvec{s}) \, d\Omega,
    \label{eq:emission_absorption}
\end{equation}
where $\kappa_\eta$ is the spectral absorption coefficient, $I_{b,\eta}$ is the spectral blackbody intensity given by Planck's law, and $G_\eta(\myvec{r})$ is the spectral incident radiation. To determine these radiative field variables, one must solve the RTE. The RTE describes the conservation of radiant energy along a direction $\myvec{s}$ for a specific wavenumber $\eta$. For a non-scattering, absorbing-emitting medium, the spectral RTE is given by
\begin{equation}
    \frac{dI_\eta(\myvec{r}, \myvec{s})}{ds} = \kappa_\eta(\myvec{r}) \left[ I_{b,\eta}(\myvec{r}) - I_\eta(\myvec{r}, \myvec{s}) \right].
    \label{eq:RTE_spectral}
\end{equation}
For opaque solid surfaces, the spectral intensity leaving the wall $I_\eta(\myvec{r}_w, \myvec{s})$ into the fluid domain ($\myvec{s} \cdot \myvec{n}_w > 0$) is the sum of emitted and reflected radiation
\begin{equation*}
    I_\eta(\myvec{r}_w, \myvec{s}) = \varepsilon_{w,\eta} I_{b,\eta}(T_w) + \frac{1-\varepsilon_{w,\eta}}{\pi} \int_{\myvec{s}' \cdot \myvec{n}_w < 0} I_\eta(\myvec{r}_w, \myvec{s}') |\myvec{s}' \cdot \myvec{n}_w| \, d\Omega',
\end{equation*}
where $\myvec{n}_w$ is the unit normal vector pointing into the domain and $\varepsilon_{w,\eta}$ is the spectral wall emissivity. For open boundaries, radiation is allowed to escape freely, and incoming rays are defined by the ambient spectral blackbody intensity
\begin{equation*}
    I_\eta(\myvec{r}_b, \myvec{s}) = I_{b,\eta}(T_{\infty}) \quad \text{for } \myvec{s} \cdot \myvec{n}_b < 0.
\end{equation*}

The primary challenge in fire simulation is the highly non-linear spectral dependence of combustion gases (e.g., CO$_2$, H$_2$O) and soot. The absorption coefficient $\kappa_\eta$ varies. While line-by-line (LBL) methods~\cite{modest2021radiative} solve Eq.~\eqref{eq:RTE_spectral} explicitly for all $\eta$, they are prohibitively expensive for CFD. In practical fire simulations, explicitly solving Eq.~\eqref{eq:RTE_spectral} for millions of wavenumbers (LBL integration) is computationally prohibitive. Therefore, two primary strategies are employed to simplify the problem. The first is the fixed radiant fraction model, where participating-medium absorption and emission are not resolved explicitly, and a constant fraction of the local heat release rate is prescribed as a radiative loss term. While efficient, this model is valid only in optically thin limits. The second involves modeling the spectral dependence of $\kappa_\eta$. Popular approaches include the weighted sum of gray gases (WSGG) model and the full-spectrum $k$-distribution (FSK) method. These methods approximate the spectral integration by solving a weighted summation of a few gray-gas RTEs. Alternatively, the simplest approximation is the Planck-mean model, which treats the medium as a single ``gray gas'' with an averaged absorption coefficient $\kappa$, assuming properties are independent of wavenumber. In this limit, Eq.~\eqref{eq:RTE_spectral} reduces to a single integro-differential equation for the total intensity $I$. By defining the effective intensity $I = \int I_\eta d\eta$ and the Planck-mean absorption coefficient $\kappa_P = (\int \kappa_\eta I_{b,\eta} d\eta) / (\sigma T^4 / \pi)$, we recover the reduced gray gas form:
\begin{equation}
    \frac{dI(\myvec{r}, \myvec{s})}{ds} = \kappa_P(\myvec{r}) \left[ I_b(\myvec{r}) - I(\myvec{r}, \myvec{s}) \right]
    \label{eq:RTE_gray}.
\end{equation}

The second major challenge lies in the numerical solution of the integro-differential RTE of Eq.~\eqref{eq:RTE_gray}. Standard numerical methods include the Monte Carlo ray tracing (MCRT) method and deterministic methods such as the DOM and the P1 approximation. While MCRT can handle arbitrary geometric complexity, it suffers from statistical noise and slow convergence. DOM (commonly used in FireFOAM) discretizes the angular space into a finite set of directions. The RTE is then solved along each discrete direction, and the angular integral is approximated using numerical quadrature, converting the original integro-differential RTE into a coupled system of linear differential equations. DOM is computationally expensive, as it requires iterative solutions over spatial meshes and angular discretizations at every CFD time step. 

From the perspective of mathematical operators, the solution of the RTE can be formalized as a mapping between physical property fields and the radiative state. The input functions to the RTE are the spatially varying absorption coefficient, $\kappa(\mathbf{r})$, and the thermodynamic temperature field $T(\mathbf{r})$, which determines the emission field via $I_b$. The output function sought is the radiation intensity, $I(\mathbf{r}, \mathbf{s})$. Derived quantities such as the incident radiation $G(\mathbf{r})$ and the radiative heat flux divergence are subsequently evaluated through numerical integration over the unit sphere. The primary goal of this work is to validate neural operator architectures for accelerated fire modeling. By integrating these surrogates into CFD workflows, we replace the direct numerical integration of the RTE with surrogate inference.

\section{Methods}
\label{sec:methods}

We begin by presenting the operator-learning problem setup, which formulates the RTE solution as a mapping between physical fields and the radiative state. We then introduce the multiple-input neural operators, followed by the nested Fourier-MIONet architecture and the datasets used for training and evaluation.

\subsection{Problem setup}

We formulate the solution of the RTE as a supervised operator learning problem. The task is to approximate the nonlinear operator $\mathcal{G}: \mathcal{U} \to \mathcal{V}$ that maps the input fields (absorption and emission properties) to the output radiative intensity field. The input function space $\mathcal{U}$ contains the spatial distributions of the Planck-mean absorption coefficient $\kappa(\mathbf{r})$ and the blackbody emission source $I_b(\mathbf{r})$ (derived from gas temperature $T$). The output function space $\mathcal{V}$ contains the angular radiation intensity $I(\mathbf{r}, \mathbf{s})$. The operator is defined as
\begin{equation*}
    \mathcal{G}: [\kappa(\mathbf{r}), T(\mathbf{r})] \mapsto I(\mathbf{r}, \mathbf{s}).
\end{equation*}
Our objective is to learn a parametric neural operator, $\mathcal{G}_\theta$, where $\theta$ denotes the trainable parameters, such that $\mathcal{G}_\theta \approx \mathcal{G}$. Since the radiative intensity depends on multiple inputs (e.g., absorption coefficient and temperature), we adopt the multiple-input operator-learning architecture.

\subsection{Multiple-input neural operators}
\label{sec:fourier-mionet}

Since the solution depends on multiple functional inputs, we employ MIONet~\cite{jin2022mionet} as a baseline architecture, which has been proposed to learn nonlinear operators with multiple input Banach spaces. We denote $\mathcal{U}_1$ and $\mathcal{U}_2$ as the Banach spaces for the absorption coefficient $\kappa(\mathbf{r})$ and the temperature $T(\mathbf{r})$, respectively, defined on a domain $D \subset \mathbb{R}^d$. We aim to learn the operator $\mathcal{G}: \mathcal{U}_1 \times \mathcal{U}_2 \to \mathcal{V}$, which maps the inputs $(\kappa, T) \mapsto I$, where $\mathcal{V}$ is the space of radiative intensity $I(\mathbf{r}, \mathbf{s})$. The input functions $\kappa \in \mathcal{U}_1$ and $T \in \mathcal{U}_2$ are discretized at a fixed set of $m$ sensor locations $\{\mathbf{r}_i\}_{i=1}^m \subset D$, yielding the vectors $\boldsymbol{\kappa} = [\kappa(\mathbf{r}_1), \dots, \kappa(\mathbf{r}_m)]^\top \in \mathbb{R}^m$ and $\mathbf{T} = [T(\mathbf{r}_1), \dots, T(\mathbf{r}_m)]^\top \in \mathbb{R}^m$. We use two neural networks (called ``branch net'') to encode the discretized inputs ($\kappa$, $T$) and one neural network (called ``trunk net'') for ($\mathbf{r}, \mathbf{s}$). The output radiative intensity $I(\mathbf{r}, \mathbf{s}) = \mathcal{G}(\kappa, T)(\mathbf{r}, \mathbf{s})$ is computed as 
$$\mathcal{G}_{\theta}(\kappa, T)(\mathbf{r}, \mathbf{s}) = \sum_{k=1}^{p} \underbrace{b_k(\kappa(\mathbf{r}_1), \kappa(\mathbf{r}_2), \ldots, \kappa(\mathbf{r}_m))}_{\text{branch}_1} \times \underbrace{c_k(T(\mathbf{r}_1), T(\mathbf{r}_2), \ldots, T(\mathbf{r}_m))}_{\text{branch}_2}  \times \underbrace{t_k(\mathbf{r},\mathbf{s})}_{\text{trunk}} + b_0,$$
where $b_0 \in \mathbb{R}$ is a trainable bias, $\{b_1, b_2, \ldots, b_p\}$ and $\{c_1, c_2, \ldots, c_p\}$ are the $p$ outputs of the branch nets, and $\{t_1, t_2,\ldots,t_p\}$ are the $p$ outputs of the trunk net. Here, $\mathcal{G}_\theta$ denotes the MIONet approximation of $\mathcal{G}$, where $\theta$ denotes the trainable parameters of the network. Note that prior to training, the inputs are scaled to ensure stable convergence. 

Vanilla MIONet architectures can face practical challenges due to the increasing complexity of data required to accurately represent the high-dimensional solution space. First, the discretized input fields contain tens to hundreds of thousands of data points, leading to high-dimensional branch inputs. We therefore employ PCA for dimension reduction and a low-dimensional representation of the functional inputs. Second, solution fields exhibit high-frequency features and sharp gradients, primarily due to the spectral bias of conventional MLPs. We replace the MLP trunk net with KAN~\cite{liu2024kan}, which uses learnable activation functions on edges, offering superior approximation capabilities for complex functions. Respectively, we propose two extensions to the MIONet: (1) PCA-MIONet processes inputs $\kappa$ and $T$ for the branch network by projecting them onto their first $N_{pc}$ principal components; (2) PCA-MIONet-KAN further enhances PCA-MIONet by replacing the standard feed-forward layers of the trunk network with KAN layers~\cite{liu2024kan, li2024kolmogorovarnold} to better capture the sharp gradients in the solution's spatial and angular dependence. The mathematical formulations for these extensions are detailed in Appendix~\ref{PCA_KAN_net}.

\begin{figure}[htbp]
    \centering
    \includegraphics[width=0.8\textwidth]{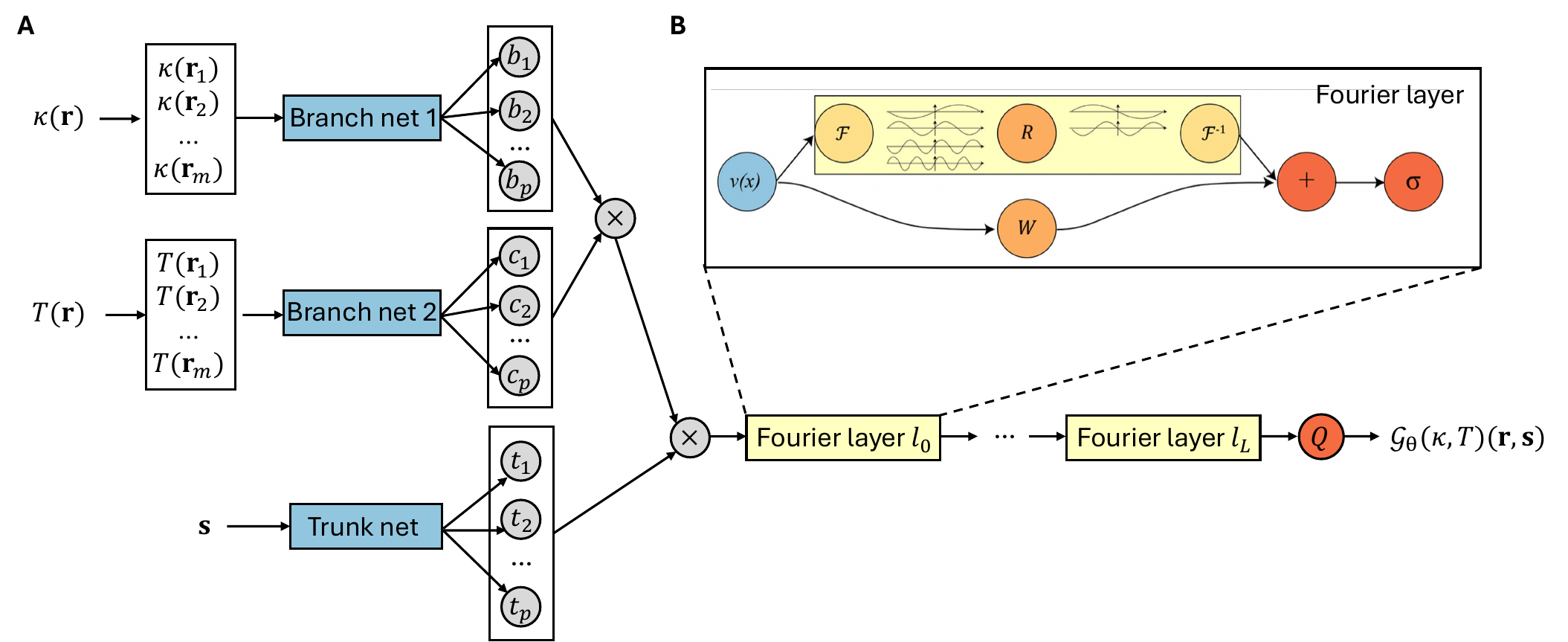}
    \caption{\textbf{Fourier-MIONet architecture.}
    (\textbf{A}) Two independent branch nets take the discretized functions $\boldsymbol{\kappa}(\mathbf{r})$ and $\mathbf{T}(\mathbf{r})$ as inputs and output $\{b_k\}_{k=1}^p$ and $\{c_k\}_{k=1}^p$, respectively. The trunk net takes $\mathbf{s}$ and outputs $\{t_k\}_{k=1}^p$.
    (\textbf{B}) The branch and trunk outputs are merged by element-wise products and then passed into $L$ Fourier layers. Each Fourier layer maps $v_j$ to $v_{j+1}$ by applying a 3D FFT $\mathcal{F}$, multiplying by learnable weights $\mathcal{R}_j$ in the Fourier space, and transforming back with $\mathcal{F}^{-1}$. A residual connection with a weight matrix $W_j$ is added.
    }
    \label{fig:fmionet}
\end{figure}

To further enhance prediction accuracy, we consider applying Fourier-MIONet~\cite{zhu2023fourier, jiang2024fourier} inspired by Fourier neural operator (FNO)~\cite{li2023fourier} (Fig.~\ref{fig:fmionet}). By integrating fast Fourier transforms into its architecture, Fourier-MIONet offers (1) higher prediction accuracy than MIONet, especially for non-smooth solutions, and (2) greater memory and computational efficiency compared to FNO~\cite{jiang2024fourier}. Specifically, we add $L$ Fourier layers after the branch-trunk merger operation (Fig.~\ref{fig:fmionet}B). Unlike the MIONet, we do not include the spatial coordinates in the trunk net, but instead they are stacked with the branch net inputs. The output of branch-trunk merger operation is lifted to a high-dimensional space and processed by a sequence of $L$ Fourier layers. The output of the ($j + 1$)th Fourier layer is computed by 
$v_{j+1} = \phi \left(\mathcal{F}^{-1}\left(\mathcal{R}_{j} \cdot \mathcal{F}(v_{j})\right) + W_j \cdot v_{j} + b_{j}\right)$, where $\phi$ is a nonlinear activation function and $b_{j}$ is a bias. For the output of the $j$-th Fourier layer $v_j$, we compute the transform using the 3D fast Fourier transform (FFT) $\mathcal{F}$ and inverse 3D FFT
$
\mathcal{F}^{-1}(\mathcal{R}_{j} \cdot \mathcal{F}(v_{j})),
$
where $\mathcal{R}_j$ is a learnable weight matrix. A residual connection with a
weight matrix $W_j$ is used. In the end, a local linear transformation $Q$ is applied by employing a shallow neural network to obtain the output $I(\mathbf{r}, \mathbf{s})$.

\subsection{Nested Fourier-MIONet}
\label{sec:nested_fourier-mionet}

\begin{figure}[htbp]
    \centering    
    \includegraphics[width=0.9\textwidth]{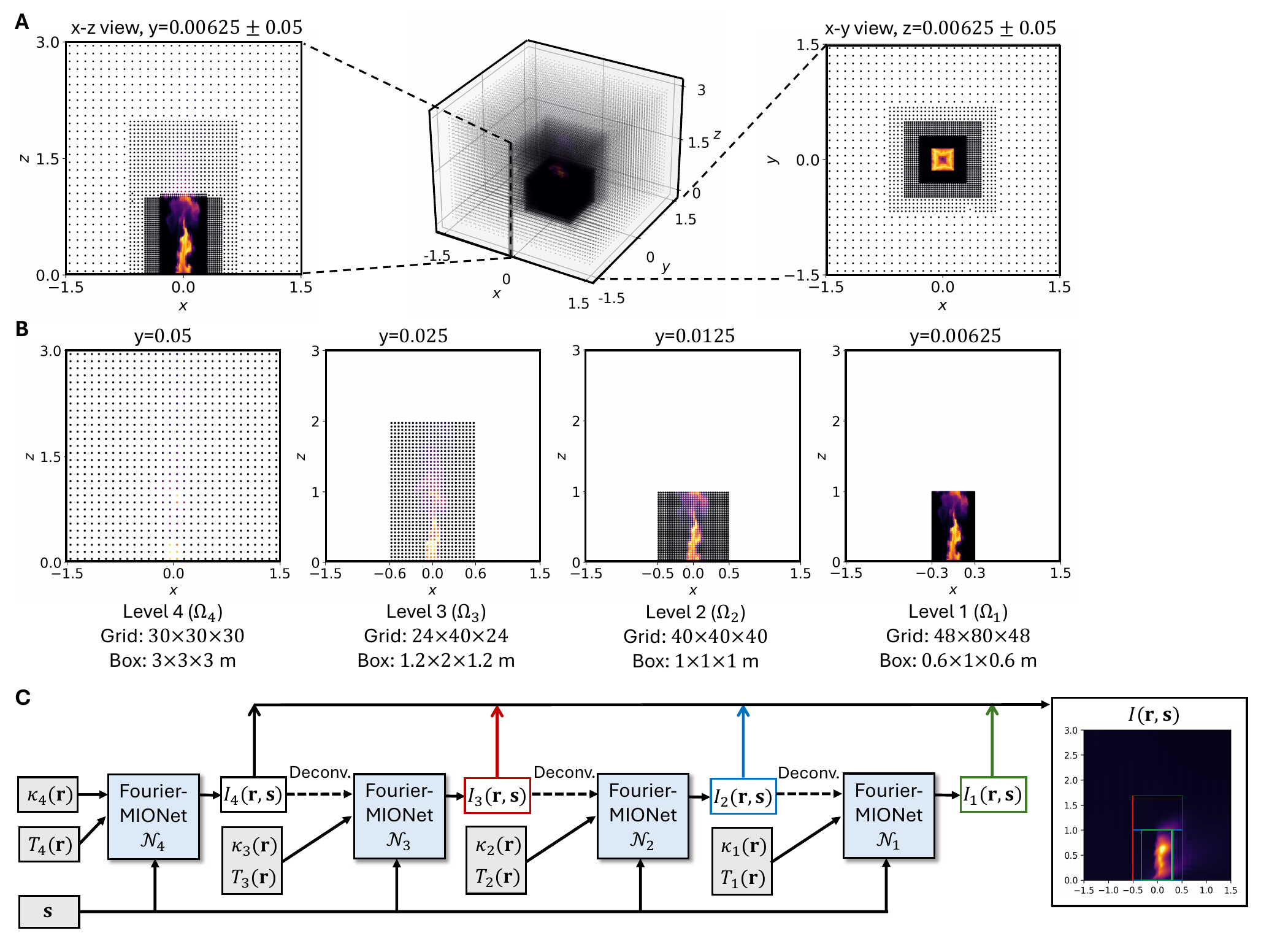}
    \caption{\textbf{Mesh refinements and nested Fourier-MIONet architecture.}
    (\textbf{A}) 2D and 3D visualizations of the mesh refinements for one example of the temperature field $T$.
    (\textbf{B}) 2D visualizations of level 4, 3, 2, and 1 refinement boxes above the burner surface with the grid resolutions and box sizes after linear interpolation.
    (\textbf{C}) Nested pipeline for predicting radiative intensity for four levels with gradually increasing resolution. Each level, starting from level 4, uses outputs from the previous coarser level on top of the absorption coefficient, temperature, and spatial directions (black arrows). The dashed arrows represent the deconvolution layers. The final global solution $I(\mathbf{r}, \mathbf{s})$ for the entire domain is assembled from the individual predictions by selecting the finest available resolution for each spatial node.
    }
    \label{fig:nfmionet}
\end{figure}

In CFD simulations, it is standard practice to employ meshes with refinements, where regions exhibiting complex dynamics are assigned finer cells while less critical areas remain coarser. For the 3D McCaffrey fire simulations, fine mesh resolution is concentrated within and around the flame/plume region to adequately resolve the entrained airflow. Consequently, a refined mesh with four levels is generated following Wang et al. (2011)~\cite{Wang11PROCI} and applied to each HRR case (see mesh refinements in Figs.~\ref{fig:nfmionet}A and B). The global domain $\Omega$ is decomposed into a hierarchy of four nested subdomains $\{\Omega_i\}_{i=1}^4$, where the index $i$ denotes the refinement level $i \in \{1, 2, 3, 4\}$. Level 4 ($\Omega_4$) represents the coarsest resolution, while Level 1 ($\Omega_1$) represents the finest. Correspondingly, we use a nested Fourier-MIONet with four different Fourier-MIONet models $\{\mathcal{N}_i\}_{i=1}^4$, where each network $\mathcal{N}_i$ approximates the radiative intensity $I_i$ on the subdomain $\Omega_i$.

%% angular directions; A dashed line

\paragraph{Training the nested Fourier-MIONets} 
For a standard, non-nested Fourier-MIONet, training is performed on a single mesh with fixed resolution using inputs $(\kappa, T)$, and inference requires only a single pass. In contrast, our nested architecture must handle multiple resolutions, with prediction at each finer level conditioned on outputs from the immediately coarser level. Accordingly, each Fourier-MIONet is trained independently at each level. The network $\mathcal{N}_4$ is trained on $\Omega_4$ using properly normalized inputs $\kappa_4, T_4$ as described in Section~\ref{sec:fourier-mionet}. For the subsequent finer levels $i \in \{3,2,1\}$, the network $\mathcal{N}_i$ is trained using its inputs $\kappa_i, T_i$ as well as the ground-truth solution $I_{i+1}$ from the coarser level $i+1$. A deconvolution (upsampling) layer is used for $I_{i+1}$ to match the resolution of $\Omega_i$. To match the resolution of $\Omega_i$ for $I_{i+1}$, we upsample $I_{i+1}$ using a learnable operator $\mathcal{U}_{i+1}$, implemented as a deconvolution operation composed of 3D convolutional layers.

\paragraph{Inference with nested prediction} 
During inference, we make a coarse-to-fine nested prediction starting from level 4 to level 1 (Fig.~\ref{fig:nfmionet}C). First, $\mathcal{N}_4$ generates the prediction $\hat{I}_4 = \mathcal{N}_4(\kappa_4, T_4)$. This prediction $\hat{I}_4$ is then upsampled via a deconvolution layer and fed as an input to $\mathcal{N}_3$ to generate $\hat{I}_3$. This process proceeds progressively in a coarse-to-fine manner, with the prediction $\hat{I}_{i+1}$ from level $i+1$ being used as input for $\mathcal{N}_i$. The prediction is defined recursively as $\hat{I}_i = \mathcal{N}_i\left(\kappa_i, T_i, \mathcal{U}_{i+1}(\hat{I}_{i+1})\right)$. The final global solution $I(\mathbf{r}, \mathbf{s})$ for the entire domain is assembled from the individual predictions $\{\hat{I}_i\}_{i=1}^4$ by selecting the finest available resolution for each spatial node:  $I(\mathbf{r}, \mathbf{s}) = \hat{I}_k(\mathbf{r}, \mathbf{s}), \text{where } k = \min \{ j \mid \mathbf{r} \in \Omega_j \}$.

\subsection{Data generation}
\label{sec:datasets}

We utilize a data-driven methodology based on CFD simulations. The CFD solver generates reference data for training and validation. FireFOAM (an OpenFOAM-based solver) is used in this work to simulate representative buoyant diffusion flames. As mentioned in Section~\ref{sec:comp_radiation_modeling}, the radiative properties are treated using the gray gas approximation (Planck-mean absorption coefficient), and the standard finite volume discrete ordinates method (fvDOM) is employed to resolve the RTE. The computational domain is defined as a rectangular enclosure. Boundary conditions are assumed to be open on all lateral and top sides to represent the ambient environment. The bottom boundary is maintained at a fixed temperature of 300 K and features a central circular burner ($0.3\times 0.3$ m$^2$) injecting gaseous methane. We generate datasets comprising the temperature field $T(\mathbf{r})$, the Planck-mean absorption coefficient field $\kappa(\mathbf{r})$, the radiative intensity $I(\mathbf{r},\mathbf{s})$, and the incident radiation $G(\mathbf{r})$. Note that our primary goal is to demonstrate MIONet as a viable surrogate model for RTE. Therefore, the chosen radiation models and numerical methods are used for this purpose rather than to provide the most physically detailed representation of the McCaffrey pool fire. Angular discretization is performed using 16 solid angles to capture the angular dependence of radiative intensity while minimizing computational overhead. Specifically, we generate three distinct datasets.

\paragraph{2D pool fire}
This dataset is derived from the standard OpenFOAM tutorial case \texttt{smallPoolFire2D}, discretized into a uniform grid of $151 \times 151$ data points. Data snapshots are collected during the developed phase from $t=35$s to $60$s, generating a total of 5,000 input-output pairs. It benchmarks the performance of different neural network architectures and hyperparameter configurations at a lower computational cost. 

\paragraph{McCaffrey pool fire of fixed fire size}
We generated five separate datasets corresponding to distinct, constant heat release rates (HRRs) ranging from $14~\mathrm{kW}$ to $58~\mathrm{kW}$, sampling at 100 Hz to capture steady-periodic puffing dynamics over the time interval $t=20$s to $60$s. CFD simulations follow the McCaffrey fire configuration \cite{Wang11PROCI} using a mesh of 272,580 cells. These datasets are employed to verify if the operator learning approach is viable for complex, turbulent 3D flames. 

\paragraph{McCaffrey pool fire of variable fire size}
A practical limitation of standard methods is that a model trained on a specific fire size typically fails if the fire grows to a new, unseen size. We generated a dataset where the fire ramps continuously from approximately $10~\mathrm{kW}$ to $60~\mathrm{kW}$. We aim to build a robust, unified model using this variable dataset to demonstrate that a single surrogate can accurately predict radiative fields for any arbitrary fire size within the operating range.

The original simulation utilizes a finite volume mesh with four levels of static local refinement, consisting of 272580 data points of cell centers (Fig.~\ref{fig:nfmionet}A). Because FVM defines cell-averaged values at the cell centers, the cell centers of one resolution level do not coincide with the subdivided cells of another resolution level, a direct transfer of values is not feasible. We therefore map the simulation results onto four structured grid levels through interpolation, providing a consistent data structure for the nested neural operator framework. To perform the interpolation efficiently, we employ linear barycentric interpolation. A Delaunay triangulation is constructed from the source mesh points, and barycentric coordinates are calculated for each target grid point within the corresponding simplex. This provides accurate interpolation for points lying inside the convex hull of the source data. For target points outside the convex hull where barycentric interpolation is undefined, we apply nearest-neighbor interpolation.
The resulting datasets span four nested domains of increasing resolution, ranging from a coarse grid of $30^3$ points over $[-1.5,1.5]\times[-1.5,1.5]\times[0,3]$ to a finest grid of $48\times80\times48$ points over $[-0.3,0.3]\times[-0.3,0.3]\times[0,1]$ (Fig.~\ref{fig:nfmionet}B). A detailed discussion of the effects of data preprocessing is provided in Appendix~\ref{appx: preprocessing}, where we show that nearest-neighbor interpolation is an efficient alternative with acceptable performance.

\subsection{Evaluation}

Model performance is assessed using the mean $L^2$ relative error and structural similarity index (SSIM)~\cite{wang2004image} for both the predicted radiative intensity and incident radiation, along with the total radiative heat loss, the radiation fraction, and energy conservation of practical importance in fire dynamics. 
The mean $L^2$ relative errors for radiative intensity $I$ and incident radiation $G$ are defined as 
$$\varepsilon_I = \frac{\| \hat{I} - I \|_2}{\| I \|_2}, \quad \varepsilon_G = \frac{\| \hat{G} - G \|_2}{\| G \|_2}.$$
The SSIMs are defined as
$$\text{SSIM}_I = \frac{(2\mu_{\hat{I}} \mu_{I} + C_1)(2\sigma_{\hat{I}I} + C_2)}{(\mu_{\hat{I}}^2 + \mu_{I}^2 + C_1)(\sigma_{\hat{I}}^2 + \sigma_{I}^2 + C_2)}, \quad \text{SSIM}_G = \frac{(2\mu_{\hat{G}} \mu_{G} + C_1)(2\sigma_{\hat{G}G} + C_2)}{(\mu_{\hat{G}}^2 + \mu_{G}^2 + C_1)(\sigma_{\hat{G}}^2 + \sigma_{G}^2 + C_2)},$$ 
where $\mu$ and $\sigma^2$ denote mean and variance, $\sigma_{\hat{I}I}$ and $\sigma_{\hat{G}G}$ are covariances, and $C_1, C_2$ are stabilization constants. We use the 2D and 3D SSIM as a perceptual metric of the similarity between two objects. Two objects are more similar when SSIM is closer to 1. 

For nested Fourier-MIONets described in Section~\ref{sec:nested_fourier-mionet}, we first evaluate the performance of each of the four Fourier-MIONets on the test data corresponding to its specific refinement level using the mean $L^2$ relative errors for radiative intensity and incident radiation and SSIMs. Once the four Fourier-MIONets are trained, global predictions are made sequentially from level 4 to level 1, where the output of one level serves as the input for the next one with convolution and deconvolution layers being used, culminating in the final global prediction. The individual error of the global predictions of each level is denoted as $\varepsilon^{\text{level}}_I$. The overall global accuracy is evaluated using relative $L^2$ errors that aggregate across all levels and test cases. For radiative intensity,
$$
\varepsilon^{\text{global}}_I = \frac{1}{N_{\text{test}} N_s} \sum_{s=1}^{N_s} \sum_{n=1}^{N_{\text{test}}} 
\frac{\sqrt{\sum_i |\hat{I}_{s,n,i} - I_{s,n,i}|^2 dV_i}}
{\sqrt{\sum_i |I_{s,n,i}|^2 dV_i}},
$$
and similarly, for incident radiation,
$$
\varepsilon^{\text{global}}_G = \frac{1}{N_{\text{test}}} \sum_{n=1}^{N_{\text{test}}} 
\frac{\sqrt{\sum_i |\hat{G}_{n,i} - G_{n,i}|^2 dV_i}}
{\sqrt{\sum_i |G_{n,i}|^2 dV_i}},
$$
where $\hat{I}_{s,n,i}$ and $I_{s,n,i}$ denote predicted and reference radiative intensities for solid angle $s$, test case $n$, and spatial element $i$, while $\hat{G}_{n,i}$ and $G_{n,i}$ denote predicted and reference incident radiation. Here $dV_i$ represents the volume of the $i$th element, $N_s$ the number of solid angles, and $N_{\text{test}}$ the number of test cases.

While a low $L_2$ relative error indicates that the model predicts radiation intensity accurately at individual data points, it does not guarantee that the global physics of the fire are preserved. Therefore, in addition to these error metrics, we also evaluate model performance by evaluating the relative errors of the total radiative heat loss, radiation fraction, and energy conservation. These quantities provide physically interpretable measures of energy exchange and serve as additional checks on the model’s accuracy. The detailed derivation and implementation are in Appendix~\ref{appx: quantities}.
The total radiative heat loss $\dot{Q}_{out}$ quantifies the net radiative flux leaving the computational domain
$$
\dot{Q}_{out} = \int_{V} (\nabla \cdot \myvec{q}_{rad})dV = \oint_{S} q_n dS = \oint_{S} (\myvec{q}_{rad} \cdot \myvec{\hat{n}}) dS.
$$
Accurately predicting this integral is crucial for assessing the thermal impact of the fire on surrounding structures and for setting correct boundary conditions in coupled simulations.
The radiation fraction is the ratio of the radiative loss to the chemical reaction rate:
$$
    \chi_R = \frac{\dot{Q}_{ems} - \dot{Q}_{abs}}{\dot{Q}_{chem}},
$$
where $\dot{Q}_{ems} = \int_V 4\pi \kappa I_b dV$, and $\dot{Q}_{abs} = \int_V \kappa \left(\int_{4\pi} I d\Omega \right) dV$.
The energy conservation is
$$\underbrace{\oint_{S} (\myvec{q}_{rad} \cdot \myvec{\hat{n}}) dS}_{\text{Net radiative heat loss}} = \underbrace{\int_{V} (\nabla \cdot \myvec{q}_{rad})dV}_{\text{Net radiative power source}}.$$ We define the absolute energy conservation residual to be $\mathcal{R}_E = \left|\oint_{S} (\myvec{q}_{rad} \cdot \myvec{\hat{n}}) dS - \int_{V} (\nabla \cdot \myvec{q}_{rad}) dV\right|$. To assess energy conservation of the predictions, we report both the ground truth $\mathcal{R}_E^{ref}$ and the predicted $\mathcal{R}_E^{pred}$ absolute energy conservation residual. 
We also report the relative errors of total radiative heat loss $\varepsilon_{\dot{Q}_{out}}$ and radiation fraction $\varepsilon_{\chi_R}$ with respect to the ground truth data, which serve as global, physically interpretable diagnostics by assessing the net radiative energy leaving the domain and the predicted radiative share of the total heat release.

\section{Results}
\label{sec:results}

In this section, we demonstrate the capabilities of our approaches using a 2D pool fire problem (Section~\ref{sec:RTEpoolfire}) and a 3D McCaffrey fire problem with fixed or variable HRRs (Section~\ref{sec:fixed_HRR} and Section~\ref{sec:variable_HRR}). For all experiments, we use the Adam optimizer for neural network training. The initial learning rate is set to  $10^{-3}$, with inverse time learning rate decay employed. The inputs are appropriately pre-processed to ensure they are on a similar scale. To enforce the non-negativity of the output radiative intensity, we apply a transformation by squaring the network output. To ensure a fair comparison, we choose model configurations with a similar number of trainable parameters. For the baseline PCA-MIONet, the two branch networks use ReLU activations~\cite{nair2010rectified} with three hidden layers of width 256, while the trunk network uses the Swish activation function~\cite{Ramachandran2017SwishAS} with three hidden layers of width 256. In addition, we employ an output-merger network with 2 hidden layers after combining the branch and trunk outputs. The PCA-MIONet-KAN replaces the MLP trunk with 3 KAN layers. The Fourier-MIONet uses four Fourier layers with 8 modes, and uses MLPs with a width of 32. The network architectures of nested Fourier-MIONets for all levels mentioned in Section~\ref{sec:nested_fourier-mionet} are shown in Appendix~\ref{appx:hyperparameters} (Table~\ref{tab:arch_nested}). The Python library DeepXDE \cite{lu2021deepxde} is utilized to implement the neural networks with the PyTorch backend. All the codes and data will be available on GitHub at \href{https://github.com/lu-group/fourier-mionet-rte}{https://github.com/lu-group/fourier-mionet-rte}.

\subsection{2D pool fire problem}
\label{sec:RTEpoolfire}

We consider the \texttt{smallPoolFire2D} case (Section~\ref{sec:datasets}). The dataset consists of 4500 samples for training and 500 for testing. We discretize the angular space into 16 directions $\mathbf{s}$ on the 2D plane. The azimuthal angles $\phi\in[0,2\pi)$ are $\phi_i=\frac{(2i+1)\Delta\phi}{2}, i=0,\ldots,15$. This 2D case features periodic fire puffing cycles, containing regions of both smooth variation and sharp gradients in the radiative intensity fields. Our goal is to systematically identify an accurate neural operator architecture for radiative transfer at a low computational cost.

\begin{table}[htbp]
    \caption{\textbf{Comparison of PCA-MIONet, PCA-MIONet-KAN, MIONet-KAN, and Fourier-MIONet on the 2D pool fire problem.}}
    \label{tab:poolfire_res}
    \centering
    \begin{tabular}{cccccc}
    \toprule
     & $\epsilon_I$ & $\text{SSIM}_I$ & $\epsilon_G$ & $\text{SSIM}_G$ & \#Params\\
    \midrule
    PCA-MIONet & 7.86\%  & 0.930 & 4.73\% & 0.978 & 1586178\\
    PCA-MIONet-KAN & 6.71\% & 0.950 & 4.16\% & 0.983 & 1576564\\
    MIONet-KAN & 7.56\% & 0.939 & 4.61\% & 0.979 & 1587232\\
    Fourier-MIONet & \textbf{3.70\%}   & \textbf{0.982} & \textbf{1.93\%} & \textbf{0.996} & 1583649\\
    \bottomrule
\end{tabular}
\end{table}

We first apply PCA-MIONet (Section~\ref{sec:fourier-mionet}), where the input field dimensionality is reduced from approximately 23K to 1.6K using PCA with 41 principal components. While this method achieves acceptable results, the error remains relatively high due to challenges in accurately capturing complex vortices and non-smooth regions in the solution fields (Table~\ref{tab:poolfire_res}). To address this, we extend the model to PCA-MIONet-KAN by incorporating KAN layers into the trunk nets, which improves accuracy and reduces error rates. We additionally evaluate MIONet-KAN without PCA.

\begin{figure}[htbp]
    \centering    
    \includegraphics[width=0.9\textwidth]{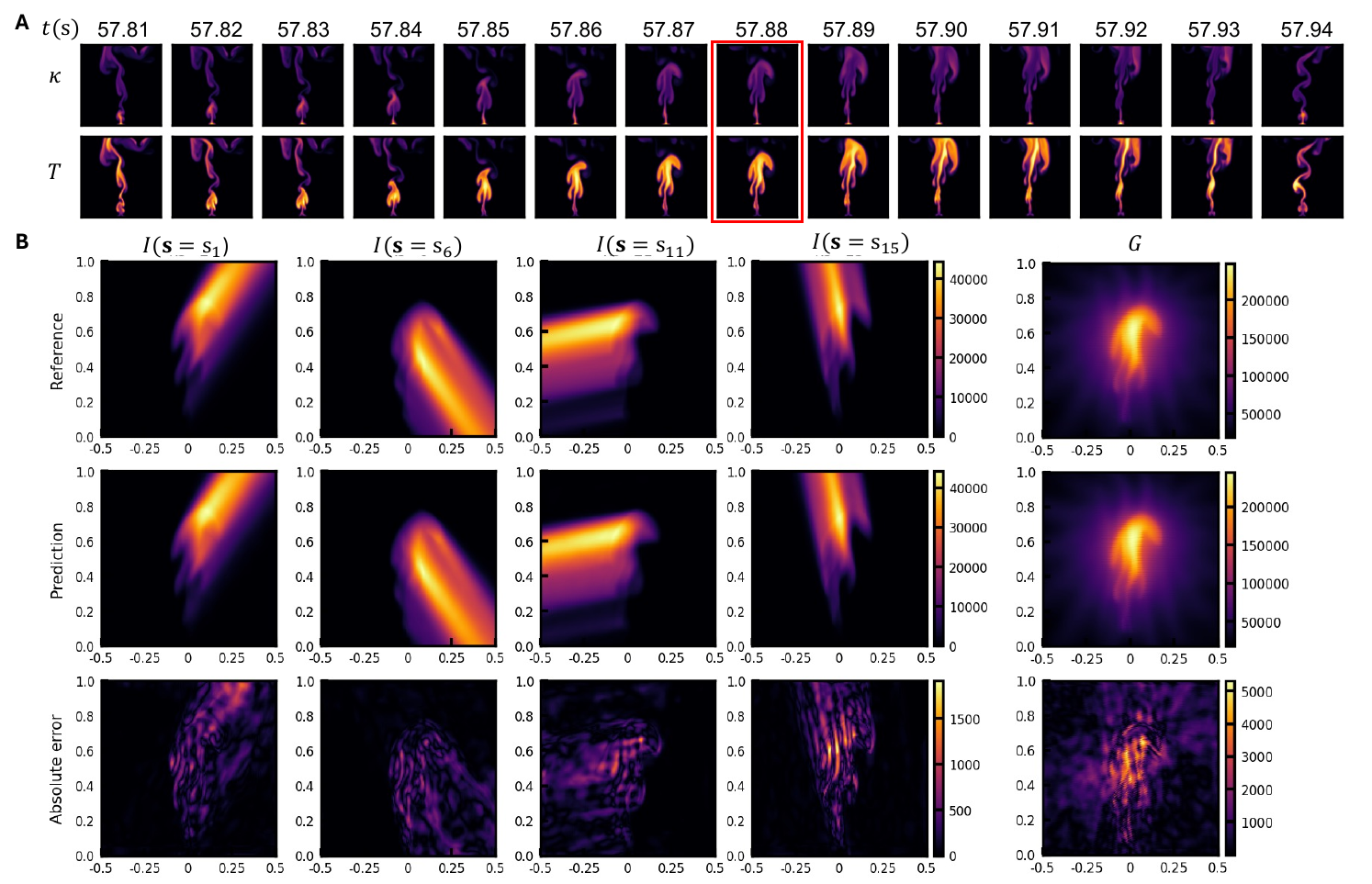}
    \caption{\textbf{Results of Fourier-MIONet for the 2D pool fire problem in Section~\ref{sec:RTEpoolfire}.}
    (\textbf{A})  Example of $\kappa$  and $T$ from the test data across half of a fire puff cycle. The red box highlights a specific example corresponding to that in (B).
    (\textbf{B}) One example of reference, prediction, and the absolute error of $I$ and $G$ on 4 different solid angles: $s_1 = (\sin{\frac{\pi}{16}}, \cos{\frac{\pi}{16}})$, $s_6 = (\sin{\frac{11\pi}{16}}, \cos{\frac{11\pi}{16}})$, $s_{11} = (\sin{\frac{21\pi}{16}}, \cos{\frac{21\pi}{16}})$, and $s_{15} = (\sin{\frac{29\pi}{16}}, \cos{\frac{29\pi}{16}})$.
    }
    \label{fig:pool_fire_res}
\end{figure}

Fourier-MIONet significantly outperforms these baselines by effectively capturing the fine-scale dynamics in the field. This approach achieves an $L^2$ relative error of 3.70\% for $I$ and $L^2$ relative error of 1.93\% for $G$, with an SSIM of 0.996 for $G$. Across the fire puff cycle (Fig.~\ref{fig:pool_fire_res}A), the Fourier-MIONet provides high-fidelity predictions of the intensity field $I$ and $G$ (see one example in Fig.~\ref{fig:pool_fire_res}B). To show whether the performance improvements are from PCA-based dimensionality reduction or from the network architecture itself, we additionally evaluate MIONet-KAN without PCA. The results show that the performance improvements mainly come from the Fourier layers. Overall, Fourier-MIONet results are significantly better than those obtained with methods without Fourier layers with a similar number of hyperparameters. This is because Fourier layers provide an explicit representation of high-frequency components, mitigating spectral bias in MLPs and making it easier to fit sharp gradients and multi-scale structures in radiative fields~\cite{rahaman2019spectral, li2020fourier}. Based on this 2D study, we adopt Fourier-MIONet as the backbone architecture for subsequent experiments on more complex settings.

\subsection{3D McCaffrey fire with fixed heat release rate}
\label{sec:fixed_HRR}

Building on the previous result, where Fourier-MIONet demonstrated superior accuracy, we now address the challenge of scaling to 3D CFD fire simulations in which the computational mesh is locally refined across multiple levels. We employ the nested Fourier-MIONet in a 3D McCaffrey fire scenario with fixed HRRs, demonstrating its ability to handle moderately large-scale problems. Following McCaffrey’s experiments~\cite{mccaffrey1979some}, we consider fires with fixed HRR values of 14, 22, 33, 45, and $58~\mathrm{kW}$. Our goal is to accurately predict the radiative intensity $I$ and relevant radiative quantities while reducing computational cost using the nested Fourier-MIONet. 

For each fixed-HRR fire case, we train the nested Fourier-MIONet on its dataset described in Section~\ref{sec:datasets} and evaluate it on the same dataset using a test split ratio of 0.03. We sample the directions $\mathbf{s}$ using a grid in the polar and azimuthal angles $(\theta, \phi)$ on a unit sphere. Here $\theta\in[0,\pi]$ is sampled with $\Delta\theta=\pi/2$, producing two points $\theta_i=\frac{(2i+1)\Delta\theta}{2},\quad i=0,1,$ and $\phi\in[0,2\pi)$ is sampled with $\Delta\phi=\pi/4$ and results in points $\phi_j=\frac{(2j+1)\Delta\phi}{2},\quad j=0,\ldots,7.$ Each pair $(\theta_i,\phi_j)$ defines one solid-angle direction $\mathbf{s}_{ij}=\bigl(\sin\theta_i\sin\phi_j, \sin\theta_i\cos\phi_j, \cos\theta_i\bigr)$, yielding $2\times 8=16$ solid angles in total. 
To understand the trade-off between accuracy and efficiency, we train four distinct models for each HRR case: Tiny, Small, Medium, and Large, which differ in network width, the number of retained Fourier modes, and the depth of Fourier layers (Section~\ref{sec:nested_fourier-mionet} and Table~\ref{tab:MIONet_config}). 
Because fires with higher HRR tend to be more turbulent and exhibit more complex flow structures that emit and absorb energy, the prediction becomes more challenging as HRR increases. To show representative examples, we focus on the lowest and highest HRR cases and report the global testing performance of the nested Fourier-MIONet for fires with $58~\mathrm{kW}$ and $14~\mathrm{kW}$ HRR. Results for other HRRs (22, 33, and $45~\mathrm{kW}$) are detailed in Appendix~\ref{appx:fixed_HRR} (Table~\ref{tab:McCaffrey_22_33_45_bottom}). The per-level $L^2$ relative errors and SSIM values for radiative intensity $I$ and incident radiation $G$ during training are reported in Tables~\ref{tab:McCaffrey_train_IG}.

\begin{figure}[htbp]
    \centering    
    \includegraphics[width=0.9\textwidth]{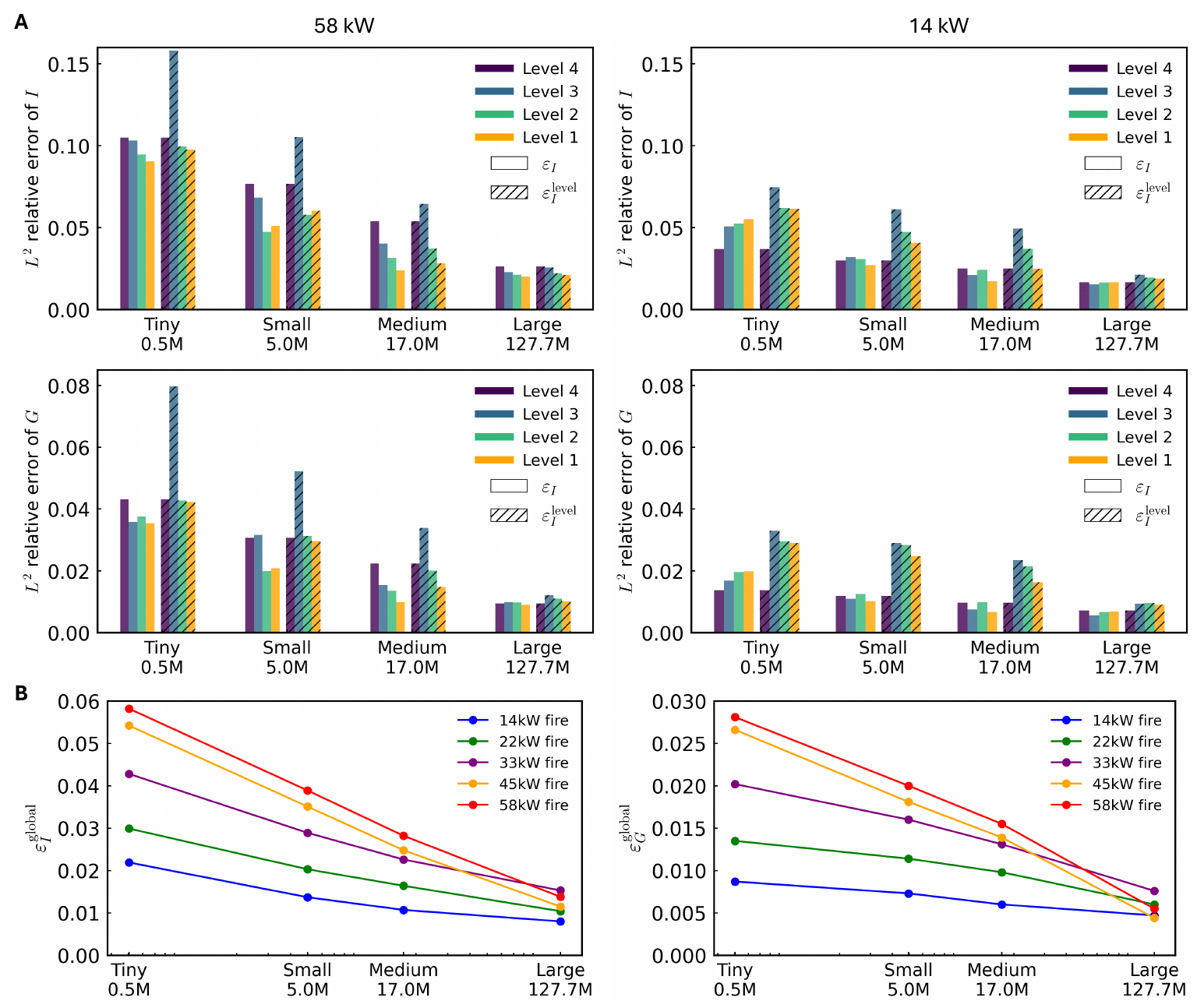
    }
    \caption{\textbf{Model accuracy across refinement levels and HRRs for the nested Fourier-MIONet.} (\textbf{A}) Per-level $L^2$ relative errors for $I$ and $G$ for the $58~\mathrm{kW}$ (left) and $14~\mathrm{kW}$ (right) McCaffrey fires. Solid bars denote the individual per-level error $\varepsilon_I$ and $\varepsilon_G$, and hatched bars denote the global inference error $\varepsilon_I^{\text{level}}$ and $\varepsilon_G^{\text{level}}$. (\textbf{B}) The $L^2$ relative errors of the radiative intensity $I$ (left) and incident radiation $G$ (right) versus the model size for 14, 22, 33, 45, and $58~\mathrm{kW}$ cases during global inference.
    }
    \label{fig:Mc_fixed_hrr_err}
\end{figure}

As expected, we observe that the errors of $58~\mathrm{kW}$ fires are generally larger than those of $14~\mathrm{kW}$ fires (Table~\ref{tab:McCaffrey_14_58}), but the nested Fourier-MIONet consistently captures these complex radiative structures. Second, increasing model capacity from tiny to large reduces prediction errors across all levels, with the large model achieving level one errors of only $2.11\%$ for $58~\mathrm{kW}$ case and $1.87\%$ for $14~\mathrm{kW}$ case for $I$ and global inference errors around $1\%$ for both cases. The SSIM values mostly achieve above 0.99, indicating high structural similarity to the reference solutions. We also note that as error accumulates through the refinement levels during global inference, the $L^2$ relative error of each level is slightly higher than that in the individual testing (Fig.~\ref{fig:Mc_fixed_hrr_err}). These results illustrate a clear trade-off between accuracy and efficiency. The tiny model is the most computationally efficient one, while the large model, which contains the highest number of parameters, achieves the best accuracy.

\begin{table*}[htbp]
  \centering
  \caption{\textbf{The global testing performance of the nested Fourier-MIONet for the 3D McCaffrey 58 kW and 14 kW fire.} We report the individual per-level errors and the global errors obtained during nested inference. The reference radiative fractions $\chi_R$ are 21.26\% and 17.83\% for the 58 kW and 14 kW fires, respectively.}
  \label{tab:McCaffrey_14_58}
  \small
  \setlength{\tabcolsep}{6.0pt}
  \begin{tabular}{c c cccc cc cccc}
    \toprule
    \multirow{2}{*}{\textbf{Fire}} &
    \multirow{2}{*}{\textbf{Model}} &
    \multicolumn{4}{c}{\textbf{Individual error $\varepsilon^{\text{level}}_I$}} &
    \multirow{2}{*}{\textbf{$\varepsilon^{\text{global}}_I$}} &
    \multirow{2}{*}{\textbf{$\varepsilon^{\text{global}}_G$}} &
    \multirow{2}{*}{$\varepsilon_{\dot{Q}_{out}}$} &
    \multirow{2}{*}{$\varepsilon_{\chi_R}$} &
    \multirow{2}{*}{$\mathcal{R}_E^{pred}$} &
    \multirow{2}{*}{$\mathcal{R}_E^{ref}$} \\
    \cmidrule(lr){3-6}
      & & Level 4 & Level 3 & Level 2 & Level 1 & & & & & & \\
    \midrule

    \multirow{4}{*}{58 kW} 
    & Tiny   & 10.50\% & 15.80\% &  9.95\% &  9.77\% & 5.82\% & 2.81\% & 0.15\% & 0.15\% & 0.14$~\mathrm{kW}$ & \multirow{4}{*}{0.01$~\mathrm{kW}$} \\
    & Small  &  7.68\% & 10.54\% &  5.78\% &  6.03\% & 3.89\% & 2.00\% & 0.04\% & 0.03\% & 0.11$~\mathrm{kW}$ & \\
    & Medium &  5.39\% &  6.45\% &  3.73\% &  2.79\% & 2.82\% & 1.55\% & 0.02\% & 0.02\% & 0.00$~\mathrm{kW}$ & \\
    & Large  &  2.63\% &  2.57\% &  2.22\% &  2.11\% & 1.38\% & 0.55\% & 0.05\% & 0.06\% & 0.01$~\mathrm{kW}$ & \\

    \midrule

    \multirow{4}{*}{14 kW}
    & Tiny   &  3.69\% &  7.47\% &  6.20\% &  6.16\% & 2.19\% & 0.87\% & 0.19\% & 0.19\% & 0.35$~\mathrm{kW}$ & \multirow{4}{*}{0.45$~\mathrm{kW}$} \\
    & Small  &  3.00\% &  6.12\% &  4.74\% &  4.07\% & 1.37\% & 0.73\% & 0.09\% & 0.09\% & 0.41$~\mathrm{kW}$ & \\
    & Medium &  2.50\% &  4.94\% &  3.71\% &  2.48\% & 1.07\% & 0.60\% & 0.05\% & 0.06\% & 0.42$~\mathrm{kW}$ & \\
    & Large  &  1.66\% &  2.12\% &  1.94\% &  1.87\% & 0.80\% & 0.47\% & 0.07\% & 0.08\% & 0.42$~\mathrm{kW}$ & \\

    \bottomrule
  \end{tabular}
\end{table*}

\begin{figure}[htbp]
    \centering
    \includegraphics[width=0.85\textwidth]{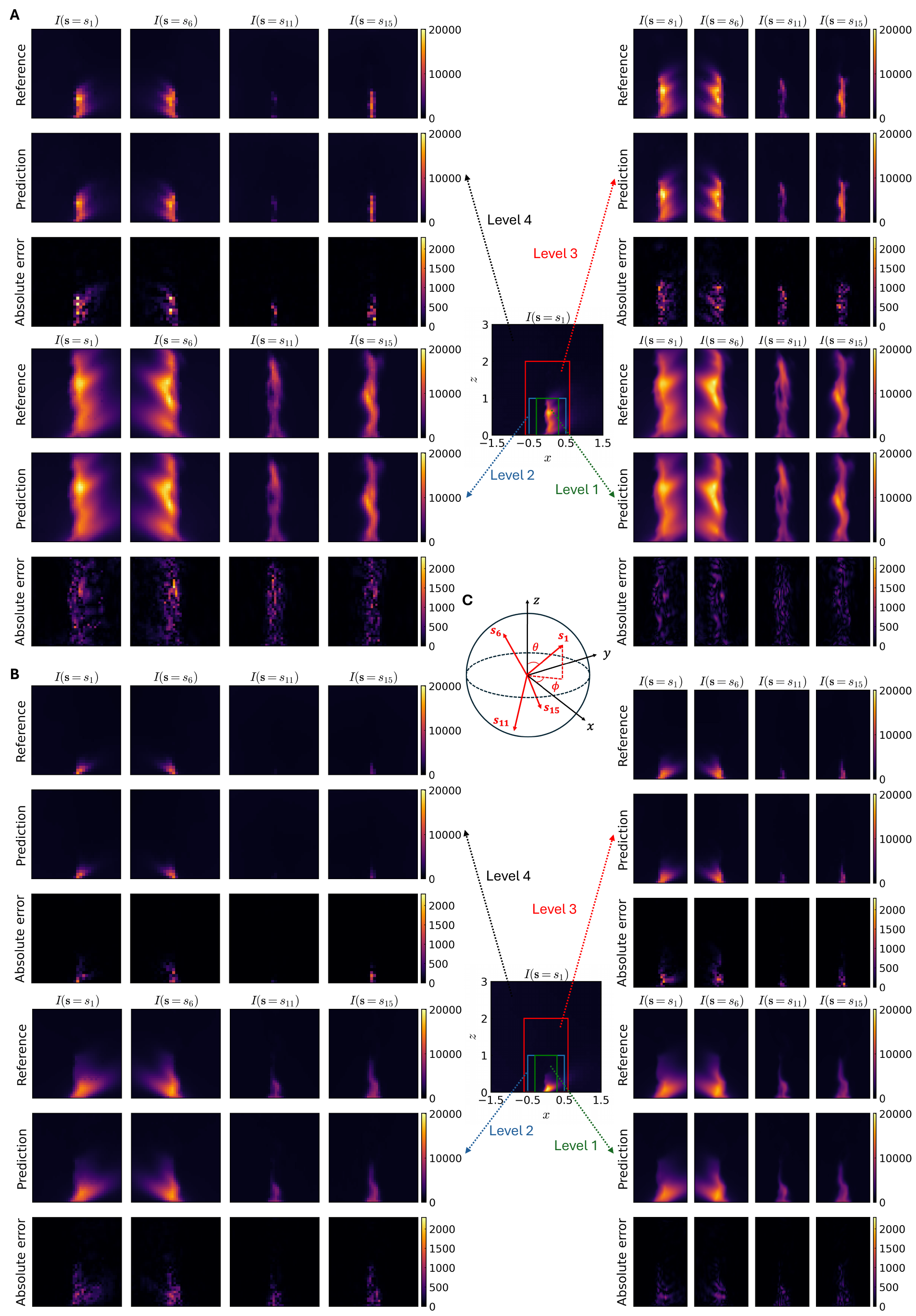}
    \caption{\textbf{Results of McCaffrey fire with HRR 58 kW and 14 kW in Section~\ref{sec:fixed_HRR} using the fixed-HRR medium model.} Examples of predictions, references, and the corresponding absolute errors of level 4, level 3, level 2, and level 1 of fire (\textbf{A}) $58~\mathrm{kW}$ and (\textbf{B}) $14~\mathrm{kW}$. The predictions are combined to obtain the final global prediction $I(\mathbf{s} = s_1)$ in the middle column. (\textbf{C}) Schematic of four representative directions from the 16 directions we sampled on the unit sphere. They are defined by the corresponding polar and azimuthal angles $s_k = (\theta_i,\phi_j)$, with $k=8i+j$, where $\theta_i=\frac{(2i+1)\Delta\theta}{2},\ i=0,1$, and $\phi_j=\frac{(2j+1)\Delta\phi}{2},\ j=0,\ldots,7$.
    }
    \label{fig:McCaffrey_fire_fixed}
\end{figure}

\begin{figure}[htbp]
    \centering    
    \includegraphics[width=0.8\textwidth]{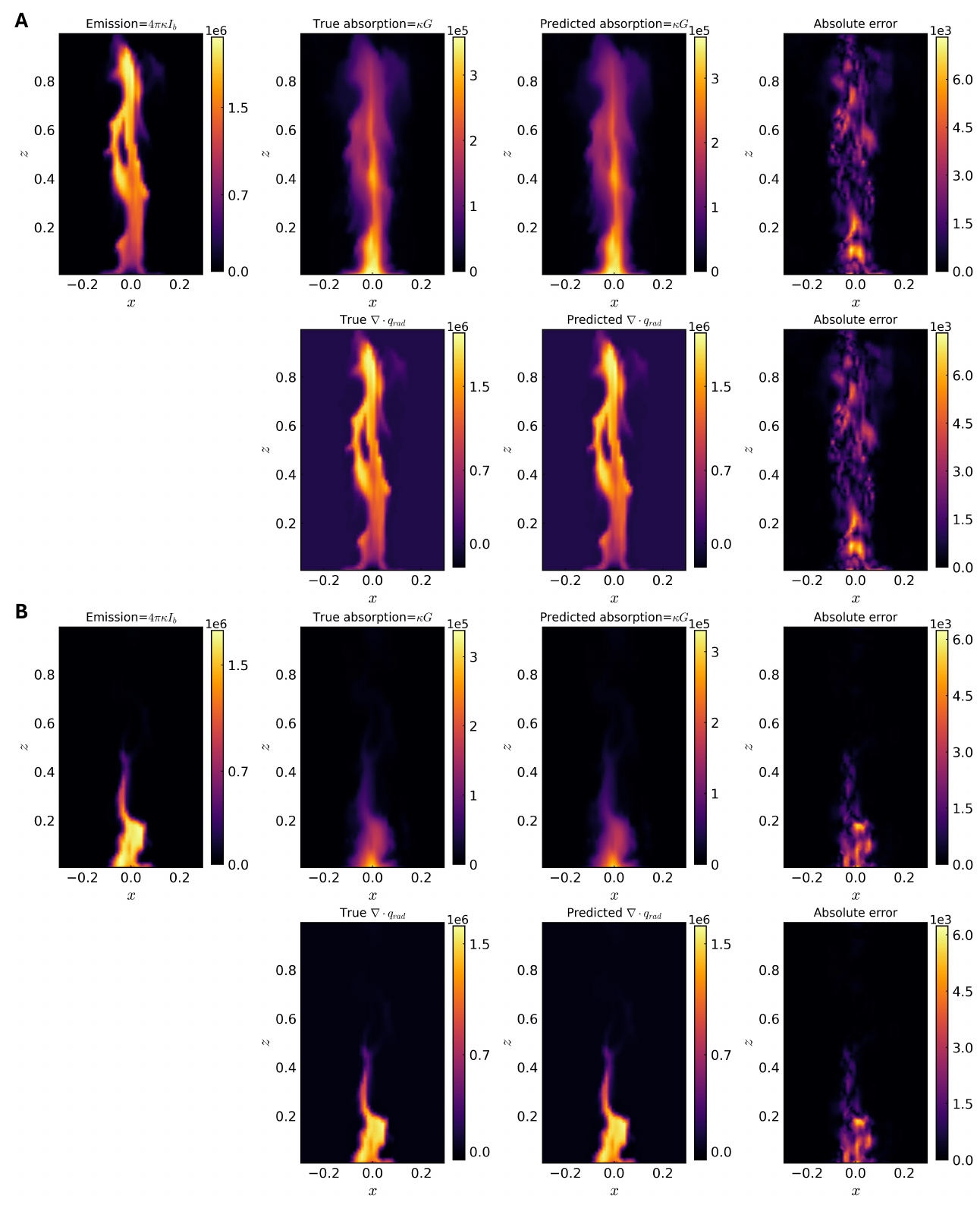}
    \caption{\textbf{Emission, absorption, and the radiative heat loss $\nabla \cdot q_{\mathrm{rad}}$ fields of the McCaffrey fire predicted from the fixed-HRR medium model.} Examples of $x$--$z$ views of predictions, references, and the corresponding absolute errors for fire with HRR (\textbf{A}) $58~\mathrm{kW}$, and (\textbf{B}) $14~\mathrm{kW}$ at $y$ = 0.00625.
    }
    \label{fig:McCaffrey_fire_fixed_energy}
\end{figure}

Fig.~\ref{fig:McCaffrey_fire_fixed} shows predictions, references, and absolute errors for the $14~\mathrm{kW}$ and $58~\mathrm{kW}$ fires using the medium model, with the combined global solutions shown in the middle column. In the 3D CFD simulations, we sample 16 directions $\mathbf{s}$ using a grid in the polar and azimuthal angles $(\theta, \phi)$ on the unit sphere, and we present 4 representative directions (Fig.~\ref{fig:McCaffrey_fire_fixed}C). The $14~\mathrm{kW}$ case is smoother and has smaller errors. Hotter regions near the base of the plume produce the strongest emission, resulting in larger values of emission $4\pi I_b$ and correspondingly higher absorption $\kappa G$ (Eq.~\ref{eq:emission_absorption}) in these lower locations (Fig.~\ref{fig:McCaffrey_fire_fixed_energy}). Moreover, the total radiative heat loss and radiation fraction are predicted with small relative errors, and the predicted energy conservation residual remains consistent with the ground truth. The nested Fourier-MIONet captures these patterns with low absolute error, indicating that the surrogate models reliably represent the radiative heat-loss behavior needed for CFD energy calculations.

\subsection{3D McCaffrey fire with variable heat release rates}
\label{sec:variable_HRR}

In practical fire scenarios, the heat release rate of fire is rarely constant, as the fuel mass flow rate changes over time. While fixed-HRR models trained from a single, fixed HRR dataset can perform well, the HRR is not a prescribed input parameter in CFD solvers like FireFOAM. This makes it infeasible to train a simple surrogate model for one fixed HRR. Moreover, a model trained on a specific fire size often fails when the fire grows to an unseen intensity. To overcome this, we develop a variable-HRR model that can accurately predict radiative transfer across a range of fire intensities without explicit knowledge of the HRR. 

\begin{figure}[htbp]
    \centering    
    \includegraphics[width=0.85\textwidth]{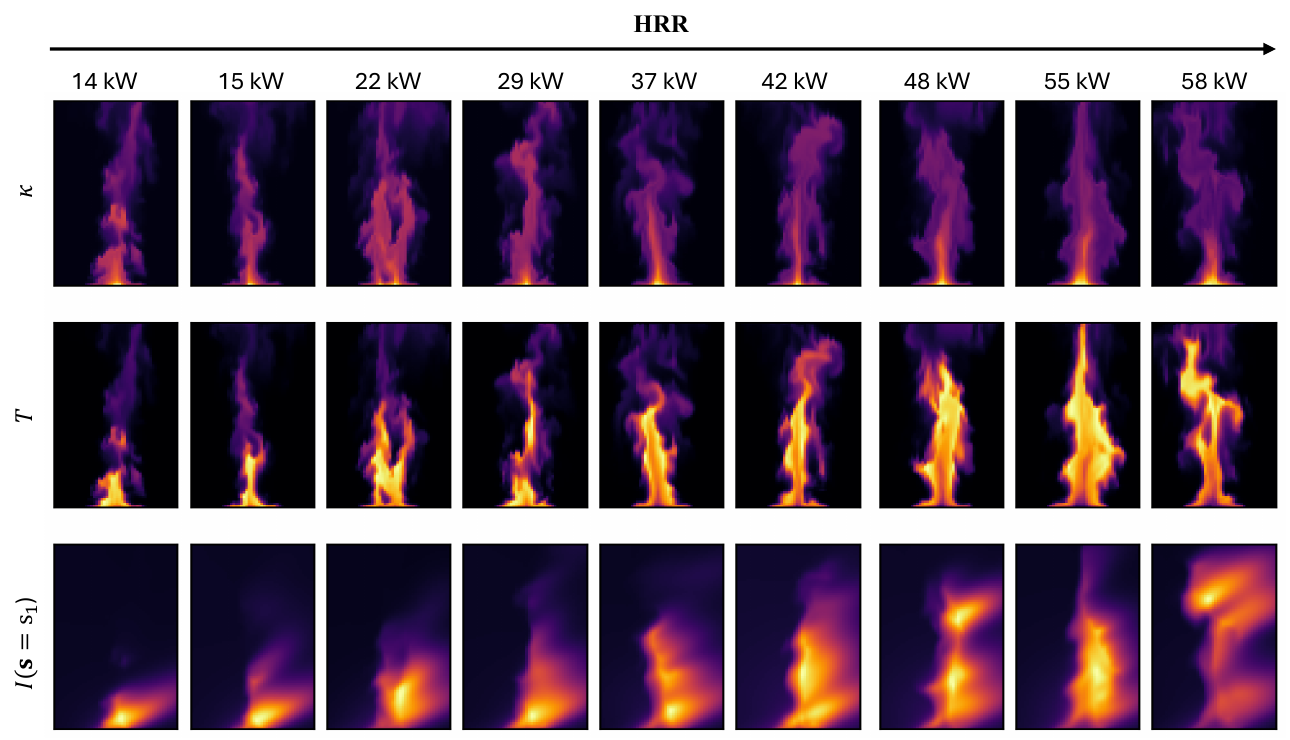}
    \caption{\textbf{Input fields with variable HRRs for the McCaffrey fires dataset in Section~\ref{sec:variable_HRR}.} 
    }
    \label{fig:Mc_increasing_hrr}
\end{figure}

To achieve this, we train the nested Fourier-MIONets on the entire dataset of the growing fire described in Section~\ref{sec:datasets}, which spans from 10$~\mathrm{kW}$ to 60$~\mathrm{kW}$. This setting is significantly more challenging than the fixed cases, as the model needs to learn to generalize across the varying radiative behaviors. As HRR increases, the fire exhibits periodic puffing but with larger flame heights, more variable plume structures, and more complex radiation intensity fields (Fig.~\ref{fig:Mc_increasing_hrr}). 

\begin{figure}[htbp]
    \centering
    \includegraphics[width=0.9\textwidth]{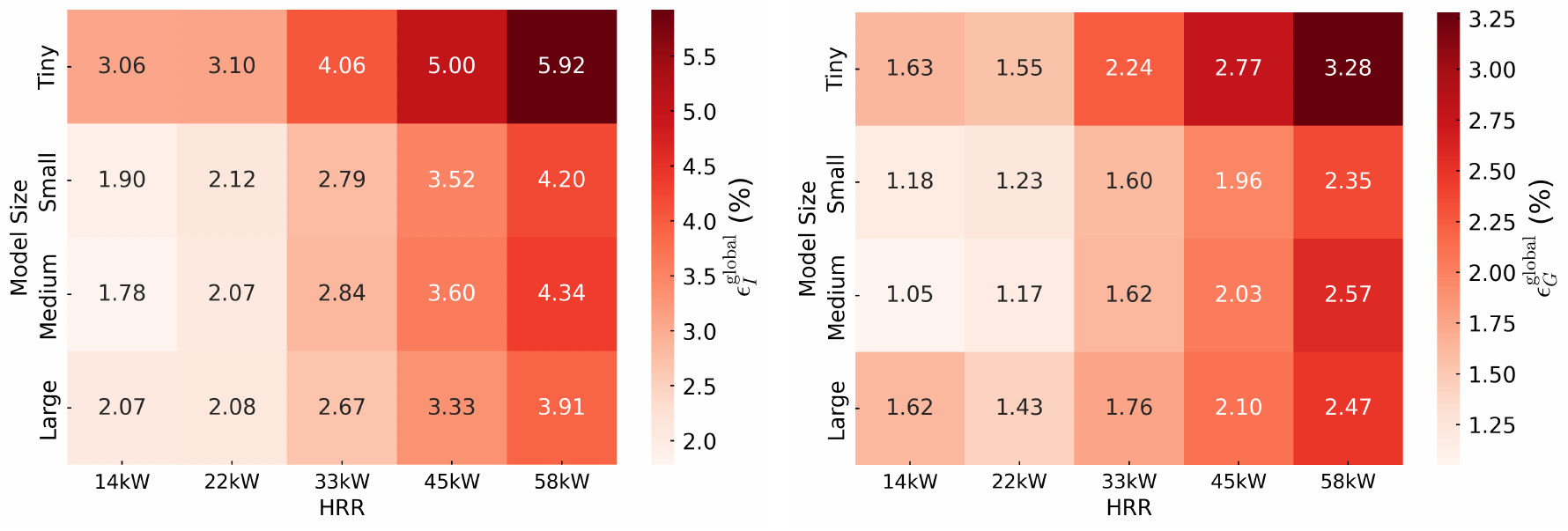
    }
    \caption{\textbf{Errors of McCaffrey fire models trained on the variable-HRR dataset and tested on fixed-HRR cases.} The global inference errors of $I$ (left) and $G$ (right) for different model sizes evaluated across fixed HRRs from 14 kW to 58 kW. Darker colors indicate larger errors.
    }
    \label{fig:Mc_all_hrr_err}
\end{figure}

Again, we train Tiny, Small, Medium, and Large models on this dataset, and we test these models on the same fixed-HRR datasets (14, 22, 33, 45, and $58~\mathrm{kW}$) from Section~\ref{sec:fixed_HRR}. Note that these fixed-HRR datasets are not included in the variable-HRR training dataset. Similarly, error increases with HRR as higher-complexity fires remain more difficult to predict. Increasing model capacity reduces prediction errors (Table~\ref{tab:McCaffrey_all_testing}, Table~\ref{tab:McCaffrey_all_training}, and Fig.~\ref{fig:Mc_all_hrr_err}). For the $58~\mathrm{kW}$ fire, the large model achieves global inference errors of $3.91\%$ for $I$ and $2.47\%$ for $G$. While this is higher than the $58~\mathrm{kW}$ fixed-HRR model's error, the relatively small increase is highly encouraging. It demonstrates that a single model can successfully learn across a wide range of fire behaviors.

\begin{table*}[htbp]
  \centering
  \caption{\textbf{3D McCaffrey fire trained on variable HRRs and tested on fixed HRR fires.} We summarize the individual per-level errors and the global errors obtained during nested inference for each HRR.}
  \label{tab:McCaffrey_all_testing}
    \begin{minipage}{\textwidth}
    \centering
    \small
    \setlength{\tabcolsep}{4.2pt}
    \begin{tabular}{c c cccc cc cccc}
        \toprule
        \multirow{2}{*}{\textbf{Model}} &
        \multirow{2}{*}{\textbf{HRR}} &
        \multicolumn{4}{c}{\textbf{Individual error $\varepsilon^{\text{level}}_I$}} &
        \multirow{2}{*}{\textbf{$\varepsilon^{\text{global}}_I$}} &
        \multirow{2}{*}{\textbf{$\varepsilon^{\text{global}}_G$}} &
        \multirow{2}{*}{$\varepsilon_{\dot{Q}_{out}}$} &
        \multirow{2}{*}{$\varepsilon_{\chi_R}$} &
        \multirow{2}{*}{$\mathcal{R}_E^{pred}$} &
        \multirow{2}{*}{$\mathcal{R}_E^{ref}$} \\
        \cmidrule(lr){3-6}
            & & Level 4 & Level 3 & Level 2 & Level 1
            & & & & & \\
        \midrule
    % ---------- Tiny ----------
        \multirow{5}{*}{Tiny}
          & $14~\mathrm{kW}$ & 4.55\% & 6.88\% & 8.85\% & 8.70\% & 3.06\% & 1.63\% & 1.80\% & 1.80\% & 0.42$~\mathrm{kW}$ & 0.45$~\mathrm{kW}$\\
          & $22~\mathrm{kW}$ & 5.69\% & 8.04\% & 7.58\% & 7.32\% & 3.10\% & 1.55\% & 0.38\% & 0.38\% & 0.63$~\mathrm{kW}$ & 0.45$~\mathrm{kW}$\\
          & $33~\mathrm{kW}$ & 7.80\% & 9.48\% & 8.38\% & 8.88\% & 4.06\% & 2.24\% & 1.19\% & 1.19\% & 0.77$~\mathrm{kW}$& 0.37$~\mathrm{kW}$\\
          & $45~\mathrm{kW}$ & 9.29\% & 10.67\% & 9.61\% & 10.27\% & 5.00\% & 2.77\% & 1.15\% & 1.15\% & 0.59$~\mathrm{kW}$& 0.15$~\mathrm{kW}$\\
          & $58~\mathrm{kW}$ & 10.40\% & 11.63\% & 10.56\% & 11.53\% & 5.92\% & 3.28\%  & 0.93\% & 0.92\% & 0.31$~\mathrm{kW}$& 0.01$~\mathrm{kW}$\\
        \midrule
    % ---------- Small ----------
        \multirow{5}{*}{Small}
          & $14~\mathrm{kW}$ & 3.48\% & 5.35\% & 4.46\% & 5.70\% & 1.90\% & 1.18\% & 1.86\% & 1.86\% & 0.31$~\mathrm{kW}$& 0.45$~\mathrm{kW}$\\
          & $22~\mathrm{kW}$ & 4.96\% & 6.45\% & 4.29\% & 4.55\% & 2.12\% & 1.23\% & 0.35\% & 0.35\% & 0.47$~\mathrm{kW}$& 0.45$~\mathrm{kW}$\\
          & $33~\mathrm{kW}$ & 7.21\% & 7.32\% & 4.67\% & 4.71\% & 2.79\% & 1.60\% & 0.21\% & 0.21\% & 0.51$~\mathrm{kW}$& 0.37$~\mathrm{kW}$\\
          & $45~\mathrm{kW}$ & 9.06\% & 8.42\% & 5.41\% & 5.22\% & 3.52\% & 1.96\% & 0.38\% & 0.38\% & 0.31$~\mathrm{kW}$& 0.15$~\mathrm{kW}$\\
          & $58~\mathrm{kW}$ & 10.39\% & 9.69\% & 6.34\% & 5.83\% & 4.20\% & 2.35\% & 0.57\% & 0.58\% & 0.24$~\mathrm{kW}$& 0.01$~\mathrm{kW}$\\
        \midrule
    % ---------- Medium ----------
        \multirow{5}{*}{Medium}
          & $14~\mathrm{kW}$ & 3.48\% & 7.10\% & 5.62\% & 3.33\% & 1.78\% & 1.05\% & 1.02\% & 1.01\% & 0.39$~\mathrm{kW}$& 0.45$~\mathrm{kW}$\\
          & $22~\mathrm{kW}$ & 4.76\% & 9.44\% & 5.91\% & 2.99\% & 2.07\% & 1.17\% & 0.04\% & 0.04\% & 0.49$~\mathrm{kW}$& 0.45$~\mathrm{kW}$\\
          & $33~\mathrm{kW}$ & 6.47\% & 11.76\% & 6.56\% & 3.76\% & 2.84\% & 1.62\% & 0.55\% & 0.55\% & 0.55$~\mathrm{kW}$& 0.37$~\mathrm{kW}$\\
          & $45~\mathrm{kW}$ & 7.85\% & 13.12\% & 7.20\% & 4.68\% & 3.60\% & 2.03\% & 1.00\% & 1.00\% & 0.49$~\mathrm{kW}$& 0.15$~\mathrm{kW}$\\
          & $58~\mathrm{kW}$ & 8.93\% & 14.16\% & 8.03\% & 5.76\% & 4.34\% & 2.57\% & 1.43\% & 1.43\% & 0.62$~\mathrm{kW}$& 0.01$~\mathrm{kW}$\\
        \midrule
    % ---------- Large ----------
        \multirow{5}{*}{Large}
          & $14~\mathrm{kW}$ & 3.70\% & 6.50\% & 5.28\% & 2.68\% & 2.07\% & 1.62\% & 1.33\% & 1.33\% & 0.35$~\mathrm{kW}$& 0.45$~\mathrm{kW}$\\
          & $22~\mathrm{kW}$ & 4.78\% & 8.30\% & 4.50\% & 1.37\% & 2.08\% & 1.43\% & 0.20\% & 0.20\% & 0.49$~\mathrm{kW}$& 0.45$~\mathrm{kW}$\\
          & $33~\mathrm{kW}$ & 6.39\% & 10.14\% & 5.18\% & 2.08\% & 2.67\% & 1.76\% & 0.70\% & 0.70\% & 0.48$~\mathrm{kW}$& 0.37$~\mathrm{kW}$\\
          & $45~\mathrm{kW}$ & 7.73\% & 11.22\% & 5.48\% & 2.92\% & 3.33\% & 2.10\% & 1.07\% & 1.07\% & 0.41$~\mathrm{kW}$& 0.15$~\mathrm{kW}$\\
          & $58~\mathrm{kW}$ & 8.59\% & 12.07\% & 6.28\% & 4.16\% & 3.91\% & 2.47\% & 1.51\% & 1.52\% & 0.60$~\mathrm{kW}$& 0.01$~\mathrm{kW}$\\
        \bottomrule
    \end{tabular}
    \end{minipage}
\end{table*}

\begin{figure}[htbp]
    \centering
    \includegraphics[width=0.85\textwidth]{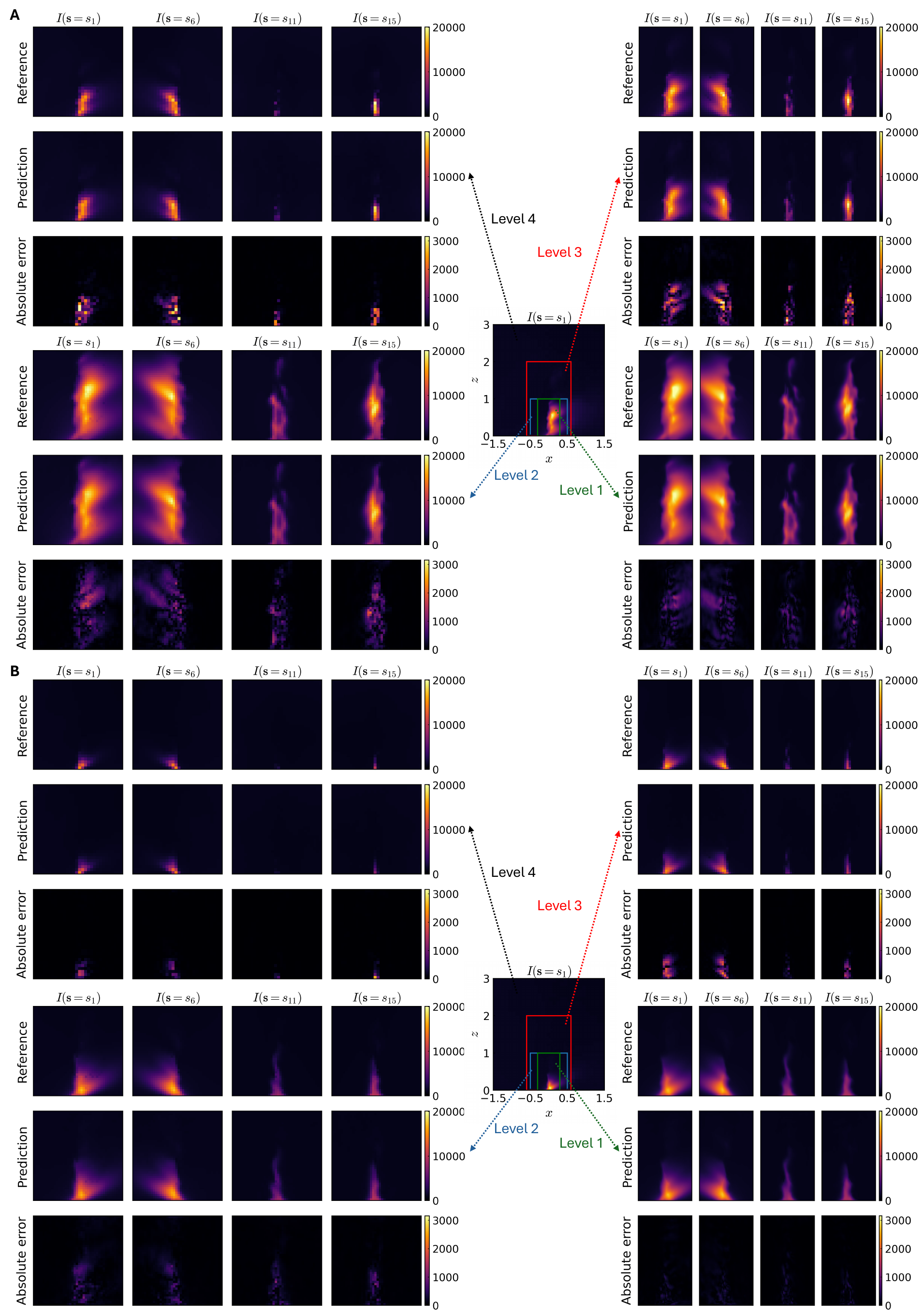}
    \caption{\textbf{Results of the McCaffrey fire using the variable-HRR medium model.} Examples of predictions, references, and the corresponding absolute errors of level 4, level 3, level 2, and level 1 of fire (\textbf{A}) $58~\mathrm{kW}$ and (\textbf{B}) $14~\mathrm{kW}$. The predictions are combined to obtain the final global prediction $I(\mathbf{s} = s_1)$ in the middle column. 
    }
    \label{fig:McCaffrey_fire_increasing}
\end{figure}

\begin{figure}[htbp]
    \centering    
    \includegraphics[width=0.8\textwidth]{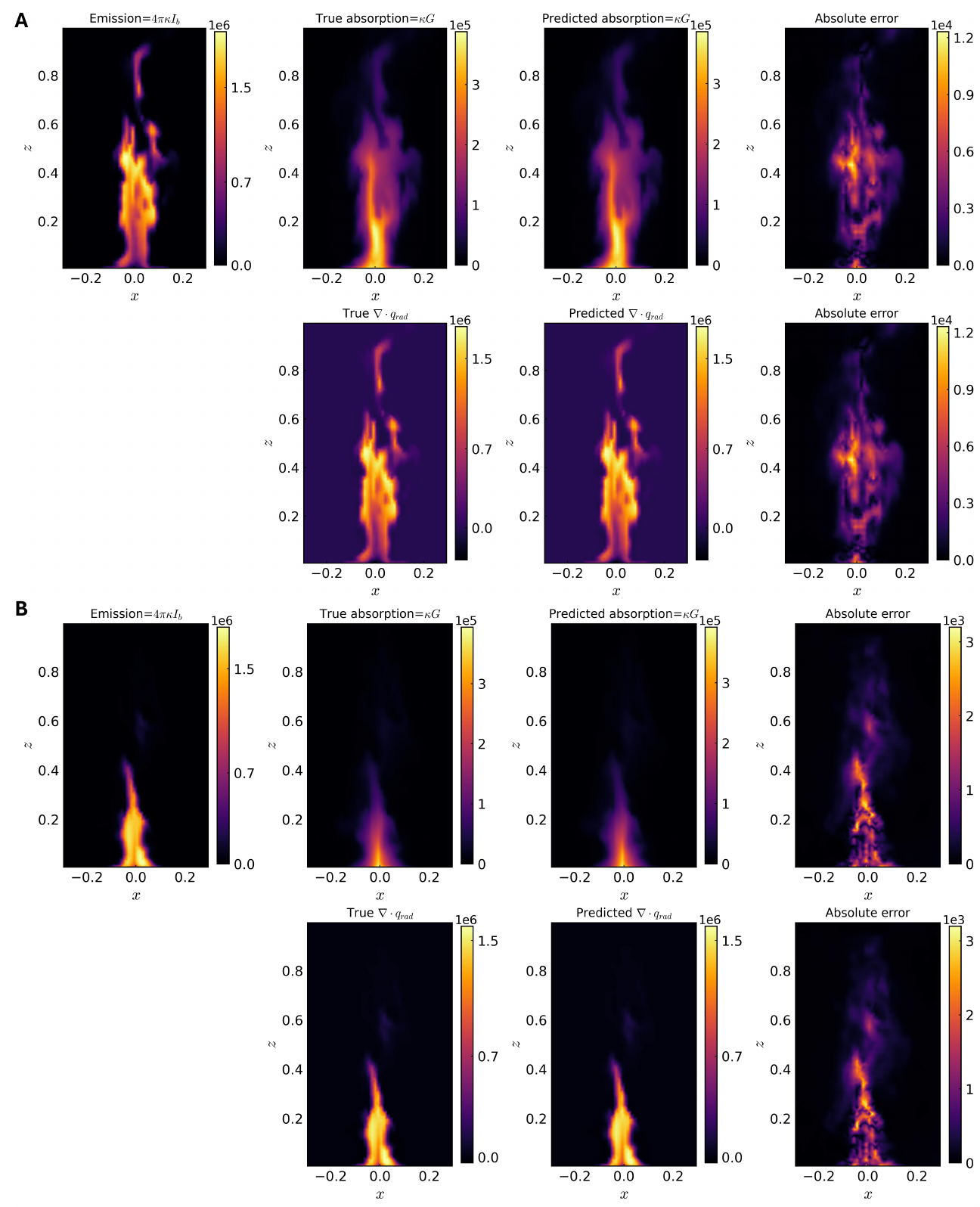}
    \caption{\textbf{Emission, absorption, and the radiative heat loss $\nabla \cdot q_{\mathrm{rad}}$ fields of the McCaffrey fire predicted from the medium variable-HRR model.} Examples of predictions, references, and the corresponding absolute errors for fire with HRR (\textbf{A}) $58~\mathrm{kW}$, and (\textbf{B}) $14~\mathrm{kW}$.
    }
    \label{fig:McCaffrey_fire_increasing_energy}
\end{figure}

The model exhibits robust adaptability, and captures both the smoother, less complex fields of the $14~\mathrm{kW}$ case while simultaneously resolving the sharp gradients of the $58~\mathrm{kW}$ fire (Fig.~\ref{fig:McCaffrey_fire_increasing}). Furthermore, the model preserves physical fidelity, correctly capturing the underlying physical structures of emission, absorption, and radiative heat loss with low absolute error (Fig.~\ref{fig:McCaffrey_fire_increasing_energy}). These results indicate that when selecting a model, the large model should be used when higher accuracy is preferred, whereas the small model offers a more computationally efficient option. More importantly, a single model can be applied to predict radiative transfer across fires of varying sizes, reducing the need for retraining and supporting practical use in CFD fire simulations.

\subsubsection{Discussion on the inference time}

Our primary motivation for developing these models is to accelerate the radiative transfer calculation, which is often the most computationally expensive component in CFD fire simulations. To quantify this acceleration, we measure the inference time for obtaining a single intensity solution across all spatial grid points and 16 solid angles using the nested Fourier-MIONet on a single NVIDIA H200 GPU.

The per-level and total inference times scale with model size. The Tiny model requires only about 0.012 seconds, while even the Large model evaluates the full multi-level solution in 0.044 seconds (Table~\ref{tab:inf_time_increasing}). For comparison, we estimate the cost of one fvDOM solve by running the same case for three iterations in FireFOAM with and without radiation. By calculating the average difference in runtime per iteration, we obtain the runtime of approximately 0.24 s, 0.14 s, and 0.10 s per fvDOM solve on 16, 32, and 48 CPU cores, respectively. While this is not a direct comparison as fvDOM is currently optimized for CPUs rather than GPUs, it highlights the potential for substantial speedups. We anticipate GPU-accelerated fvDOM implementations in FireFOAM in future work, at which point a more direct comparison will be possible. In addition, integrating the proposed surrogate models directly into the solver will be an important direction for future development.

\begin{table}[htbp]
  \centering
  \small
  \caption{\textbf{Inference time (s) of the nested Fourier-MIONet on the 3D McCaffrey fire with variable HRRs, compared with fvDOM CPU runtime.}}
  \label{tab:inf_time_increasing}
  \setlength{\tabcolsep}{4.0pt}
\begin{tabular}{c cccc cc}
    \toprule
    \multirow{2}{*}{\textbf{Model}} &
    \multicolumn{4}{c}{\textbf{Individual inference time}} &
    \multirow{2}{*}{\textbf{Total inference time}} &
    \multirow{2}{*}{\textbf{fvDOM runtime (CPU, 48 cores)}} \\
    \cmidrule(lr){2-5}
        & Level 4 & Level 3 & Level 2 & Level 1
        & & \\
    \midrule
    Tiny & 0.0027 & 0.0034 & 0.0023 & 0.0073 & 0.0157 & \multirow{4}{*}{0.10} \\
    Small & 0.0038 & 0.0021 & 0.0040 & 0.0120 & 0.0219 & \\
    Medium & 0.0039 & 0.0025 & 0.0050 & 0.0147 & 0.0261 & \\
    Large & 0.0043 & 0.0047 & 0.0094 & 0.0259 &  0.0443 & \\
    \bottomrule
  \end{tabular}
\end{table}

\section{Conclusions}

In this study, we employ Fourier-MIONets to model radiative heat transfer for both 2D and 3D fire scenarios, specifically the small pool fire and McCaffrey fire cases. The integration of PCA for dimensionality reduction and KAN layers for capturing non-smooth features enhanced the performance of MIONets, while Fourier-MIONets further improved accuracy. The nested Fourier-MIONets enabled accurate global predictions by sequentially combining results from different refinement levels. Our experiments demonstrate the effectiveness of Fourier-MIONets approaches in learning the solution operator of RTE, capturing complex features such as vortices and non-smooth regions in radiative fields. We further extend the method to generalize across McCaffrey fires with varying HRRs, and provide four models of different sizes that offer flexible trade-offs between accuracy and inference speed. 

These results establish a high-accuracy surrogate modeling framework that paves the way for a machine learning-based radiation model as a replacement for traditional radiation solvers in FireFOAM, with the goal of accelerating fully coupled radiation CFD simulations. Future work will move beyond proof-of-concept fire cases and apply these surrogates to complex, industrial-scale scenarios, such as 16-ft parallel panel tests and rack storage fires. In standard fire tests, fire sizes, boundary conditions, and radiative characteristics can differ substantially from one test to another. We will develop models that generalize robustly across a range of HRRs and investigate transfer learning strategies across materials and scales, aiming to provide a powerful tool for evaluating sprinkler performance and fire risk in complex storage environments.

\section*{Acknowledgements}

This work was supported by the FM fire modeling strategic research program and the U.S. Department of Energy Office of Advanced Scientific Computing Research under Grants No.~DE-SC0025593 and No.~DE-SC0025592. This research used resources of the National Energy Research Scientific Computing Center (NERSC), a Department of Energy User Facility (project m4946).

\section*{Declaration of competing interest}
The authors declare that they have no competing financial interests or personal relationships that could have appeared to influence the work reported in this paper.

\appendix
\renewcommand{\appendixname}{}

\input{appendix}

\bibliographystyle{unsrt}
\bibliography{main}

\end{document}

%% file: appendix.tex
\section{PCA-MIONet and PCA-MIONet-KAN}
\label{PCA_KAN_net}

We detail the mathematical formulations for the two proposed extensions to the baseline MIONet, PCA-MIONet and PCA-MIONet-KAN, as described in Section~\ref{sec:fourier-mionet}. 

\paragraph{PCA for dimensionality reduction of the inputs}
Let $\{\mathbf{r}_i\}_{i=1}^m\subset D$ be fixed sensor locations, and let $\boldsymbol{\kappa} = [\kappa(\mathbf{r}_1), \dots, \kappa(\mathbf{r}_m)]^\top \in \mathbb{R}^m$ and $\mathbf{T} = [T(\mathbf{r}_1), \dots, T(\mathbf{r}_m)]^\top \in \mathbb{R}^m$ denote the high-dimensional input vectors for the absorption coefficient and temperature, respectively. To reduce the computational cost due to high-dimensional inputs, we employ principal component analysis (PCA) to project the inputs onto a lower-dimensional linear subspace. Here, we describe the PCA process for $\boldsymbol{\kappa}$, and the procedure for $\mathbf{T}$ is the same. Given a training dataset of $N$ samples $\mathcal{D}_{\kappa} = \{\boldsymbol{\kappa}^{(n)}\}_{n=1}^N$, we first compute the empirical mean $\bar{\boldsymbol{\kappa}} = \frac{1}{N} \sum_{n=1}^{N} \boldsymbol{\kappa}^{(n)} \in \mathbb{R}^m$. We then construct the centered data matrix $\mathbf{X}_{\kappa} = [\boldsymbol{\kappa}^{(1)} - \bar{\boldsymbol{\kappa}}, \dots, \boldsymbol{\kappa}^{(N)} - \bar{\boldsymbol{\kappa}}] \in \mathbb{R}^{m \times N}$. The empirical covariance matrix $\mathbf{\Sigma}_{\kappa} \in \mathbb{R}^{m \times m}$ is given by $\mathbf{\Sigma}_{\kappa} = \frac{1}{N-1} \mathbf{X}_{\kappa} \mathbf{X}_{\kappa}^\top$. We then compute the eigenvectors $\{\boldsymbol{\psi}_j\}_{j=1}^m$ of $\mathbf{\Sigma}_{\kappa}$ corresponding to the sorted eigenvalues $\lambda_1 \ge \lambda_2 \ge \dots \ge \lambda_m \ge 0$. To reduce dimensionality, we keep the first $N_{pc} \ll m$ eigenvectors corresponding to the largest eigenvalues. We define the projection matrix $\mathbf{\Phi}_{\kappa} = [\boldsymbol{\psi}_1, \boldsymbol{\psi}_2, \dots, \boldsymbol{\psi}_{N_{pc}}] \in \mathbb{R}^{m \times N_{pc}}$. For any new input $\boldsymbol{\kappa}$, the reduced feature vector is $\tilde{\boldsymbol{\kappa}} = \mathbf{\Phi}_{\kappa}^\top (\boldsymbol{\kappa} - \bar{\boldsymbol{\kappa}}) \in \mathbb{R}^{N_{pc}}$. Applying the same procedure to $\mathbf{T}$, we have $\tilde{\mathbf{T}}\in\mathbb{R}^{N_{pc}}$.
The PCA-MIONet is then constructed by feeding these reduced representations into the branch networks: 
$
\mathcal{G}_{\theta}^{\mathrm{PCA-MIONet}}(\kappa, T)(\mathbf{r}, \mathbf{s})
=
\sum_{k=1}^{p} b_k\!\left(\tilde{\boldsymbol{\kappa}}\right)\,
c_k\!\left(\tilde{\mathbf{T}}\right)\,
t_k(\mathbf{r}, \mathbf{s})
+ b_0,
$
where $\{b_k(\cdot)\}_{k=1}^p$ and $\{c_k(\cdot)\}_{k=1}^p$ are the outputs of the two branch networks, $\{t_k(\mathbf{r},\mathbf{s})\}_{k=1}^p$ are the trunk outputs, and $b_0\in\mathbb{R}$ is a bias term.

\paragraph{PCA-MIONet-KAN enhances the trunk network with efficient KAN layers} 
To better capture sharp gradients and high-frequency features in the radiative intensity, we use PCA-MIONet-KAN. This architecture retains the PCA-reduced branch inputs $\tilde{\boldsymbol{\kappa}}$ and $\tilde{\mathbf{T}}$ but replaces the standard MLP trunk with a Kolmogorov-Arnold network (KAN)~\cite{liu2024kan}. We adopt the efficient FastKAN~\cite{li2024kolmogorovarnold}, which approximates the original B-spline basis using Gaussian radial basis functions. We use an $L$-layer FastKAN as the trunk net to encode the inputs $\mathbf{z}^{(0)} = (\mathbf{r}, \mathbf{s})$. For layer $l+1$, the output of the $j$-th node is $z_{j}^{(l+1)} = \sum_{i=1}^{n_l} \phi_{l, i, j}(z_{i}^{(l)}) + b_j^{(l+1)}, \quad j = 1, \dots, n_{l+1}$, where $n_l$ is the width of layer $l$, and $b_j^{(l+1)}$ is a bias term. The learnable activation function $\phi_{l, i, j}(x)$ is defined as $\phi(x) = w_{base} \text{SiLU}(x) + \sum_{q=1}^{Q} w_q \exp\left( - \left( \frac{x - \mu_q}{h} \right)^2 \right)$. Here, $\text{SiLU}(x)$ is a global base activation, and the latter term captures local variations using Gaussian RBFs centered at $\mu_q$ with bandwidth $h$. In this work, to ensure a comparable number of parameters with the other baseline architectures, we set the trunk depth to $L=3$, the hidden layer widths to $n_1=n_2=18$, the output width to $n_3=256$, and the number of grid points to $Q=50$. The PCA-MIONet-KAN approximation is $\mathcal{G}_{\theta}^{\mathrm{PCA-MIONet-KAN}}(\kappa, T)(\mathbf{r}, \mathbf{s})
=
\sum_{k=1}^{p} b_k\!\left(\tilde{\boldsymbol{\kappa}}\right)\,
c_k\!\left(\tilde{\mathbf{T}}\right)\,
t_k^{\mathrm{KAN}}(\mathbf{r}, \mathbf{s})
+ b_0,$ where $\{b_k(\cdot)\}_{k=1}^p$ and $\{c_k(\cdot)\}_{k=1}^p$ are the outputs of the two branch networks, $\{t_k^{\mathrm{KAN}}(\mathbf{r},\mathbf{s})\}_{k=1}^p$ are the outputs of the KAN trunk network, and $b_0\in\mathbb{R}$ is a bias term.

\section{Discussion on data preprocessing}
\label{appx: preprocessing}

\begin{figure}[htbp]
    \centering
    \includegraphics[width=0.9\textwidth]{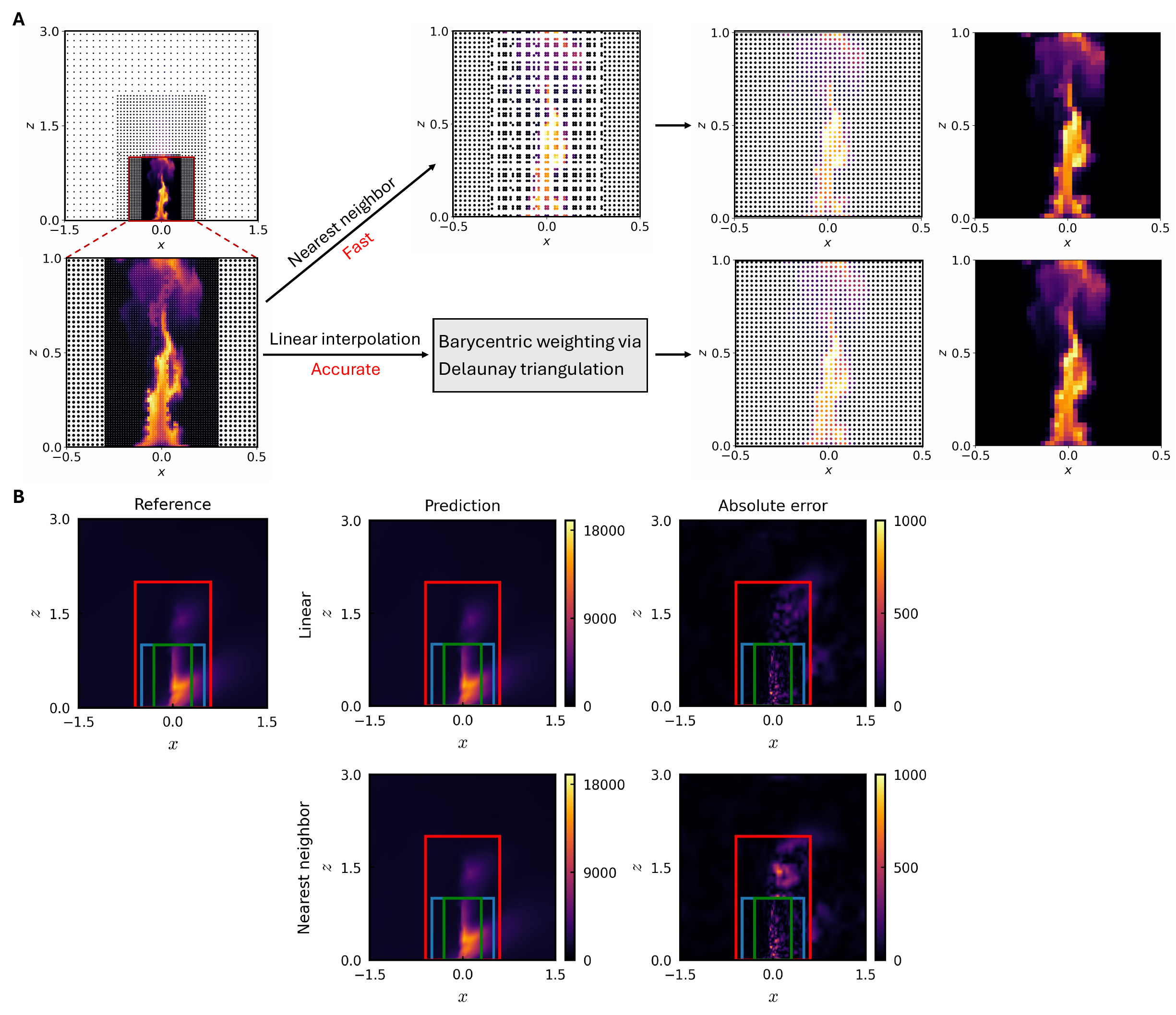}
    \caption{\textbf{Illustration and comparison of data preprocessing approaches.} (\textbf{A}) Comparison of the nearest neighbor interpolation approach (top path) and the linear interpolation approach (bottom path) on the raw mesh with a $T$ example at level 2. The nearest neighbor interpolation approach performs a direct point lookup based on minimal Euclidean distance and assigns the corresponding values to the target grid locations. The linear interpolation approach utilizes barycentric weighting via Delaunay triangulation and produces smoother and more accurate fields. (\textbf{B}) Reference, global predictions, and absolute errors of $I(\mathbf{s}=s_1)$ obtained using the two preprocessing approaches for direct comparison.
    }
    \label{fig:preprocess_comparison}
\end{figure}

As mentioned in Section~\ref{sec:datasets}, the raw FireFOAM simulation data need to be interpolated onto four structured, nested grid levels. Our primary results employ linear interpolation, which provides smoother fields and higher accuracy (Fig.~\ref{fig:preprocess_comparison}A bottom path). To facilitate easier integration into numerical solvers and reduce the computational overhead, we introduce nearest-neighbor interpolation as a simpler, faster alternative (Fig.~\ref{fig:preprocess_comparison} top path). This method eliminates the need for triangulation or barycentric weighting and requires only efficient spatial point filtering.

To assess the impact of these preprocessing choices, we trained and tested models on the $58~\mathrm{kW}$ fire using datasets generated from both interpolation methods (Table~\ref{tab:McCaffrey_interpolation}). We observe that linear interpolation results in relatively higher accuracy (Fig.~\ref{fig:preprocess_comparison}B). Nevertheless, the performance of nearest-neighbor interpolation is also acceptable for lower computational time. These experiments demonstrate that the model remains robust even under this coarser preprocessing approximation, making nearest-neighbor interpolation a practical and efficient option for coupling the surrogate models with CFD solvers such as FireFOAM.

\begin{table*}[htbp]
  \centering
  \caption{\textbf{Comparison of linear versus nearest-neighbor interpolation on 3D McCaffrey $58~\mathrm{kW}$ fire with per-level testing performance.} The top panel reports per-level $L^2$ relative errors and SSIM values for radiative intensity $I$ and incident radiation $G$ during training of the medium model. The lower panel summarizes the individual per-level errors and the global errors obtained during nested inference.}
  \label{tab:McCaffrey_interpolation}
  \setlength{\tabcolsep}{18.0pt}
  \begin{minipage}{\textwidth}
  \centering
  \small
  \begin{tabular}{c c c c c c}
    \toprule
    \multirow{2}{*}{\textbf{Interpolation}} &
    \multirow{2}{*}{\textbf{Level}} & 
    \multicolumn{2}{c}{\textbf{$I$}} &
    \multicolumn{2}{c}{\textbf{$G$}} \\ 
    \cmidrule(lr){3-4} \cmidrule(l){5-6}
       & & $\varepsilon_I$ & SSIM$_I$ & $\varepsilon_G$ & SSIM$_G$ \\
    \midrule
    \multirow{4}{*}{Linear}
      & 4 & 5.39\% & 0.995 & 2.24\% & 0.991 \\
      & 3 & 4.02\% & 0.997 & 1.54\% & 0.998 \\
      & 2 & 3.15\% & 0.997 & 1.36\% & 0.999 \\
      & 1 & 2.39\% & 0.994 & 0.99\% & 0.999 \\
      \midrule
    \multirow{4}{*}{Nearest neighbor}
      & 4 & 5.51\% & 0.995 & 2.30\% & 0.990 \\
      & 3 & 4.64\% & 0.996 & 1.69\% & 0.998 \\
      & 2 & 4.66\% & 0.934 & 1.89\% & 0.998 \\
      & 1 & 2.39\% & 0.994 & 0.99\% & 0.999 \\
    \bottomrule
\end{tabular}
  \end{minipage}

\vspace{0.1em} % vertical space between the two parts

\begin{minipage}{\textwidth}
\centering
\small
\setlength{\tabcolsep}{3.2pt}
\begin{tabular}{c cccc cc cccc}
    \toprule
    \multirow{2}{*}{\textbf{Interpolation}} &
    \multicolumn{4}{c}{\textbf{Individual error $\varepsilon^{\text{level}}_I$}} &
    \multirow{2}{*}{\textbf{$\varepsilon^{\text{global}}_I$}} &
    \multirow{2}{*}{\textbf{$\varepsilon^{\text{global}}_G$}} &
    \multirow{2}{*}{$\varepsilon_{\dot{Q}_{out}}$} &
    \multirow{2}{*}{$\varepsilon_{\chi_R}$} &
    \multirow{2}{*}{$\mathcal{R}_E^{pred}$} &
    \multirow{2}{*}{$\mathcal{R}_E^{ref}$}\\
    \cmidrule(lr){2-5}
        & Level 4 & Level 3 & Level 2 & Level 1
        & & & & & \\
    \midrule
    Linear & 5.39\% & 6.45\% & 3.73\% & 2.79\% & 2.82\% & 1.55\% & 0.02\% & 0.02\% & 0.00$~\mathrm{kW}$& 0.01$~\mathrm{kW}$\\
    \midrule
    Nearest neighbor & 5.51\% & 7.31\% & 5.97\% & 3.39\% & 3.09\% & 1.70\% & 0.39\% & 0.40\% & 0.18$~\mathrm{kW}$& 0.01$~\mathrm{kW}$\\
    \bottomrule
  \end{tabular}
\end{minipage}
\end{table*}

%% add the outputs and numbers

\section{Radiative quantities}
\label{appx: quantities}

\paragraph{Total radiative heat loss} 
The total radiative heat loss $\dot{Q}_{out}$ quantifies the net radiative flux leaving the computational domain. It is the integration over all the surface elements with surface normal $\myvec{\hat{n}}$:
\begin{equation*}
    \dot{Q}_{out} = \int_{V} (\nabla \cdot \myvec{q}_{rad})dV = \oint_{S} q_n dS = \oint_{S} (\myvec{q}_{rad} \cdot \myvec{\hat{n}}) dS 
    \approx \sum_{f \in \text{all faces}} \dot{Q}_{out, f},
\end{equation*}
where $\dot{Q}_{out, f}$ represents face $f$ out of the six boundary faces of the cubic domain in the 3D McCaffrey case. This is further approximated by summing the flux over all $N_f$ discrete surface elements on that face, each with area $\Delta S_j$: $\dot{Q}_{out, f} \approx \sum_{j}^{N_f} q_{n,j} \Delta S_j$, where $N_f$ denotes the total number of surface elements. 
Here, the radiative heat flux on a surface element with a surface normal $\myvec{\hat{n}}$ is
\begin{equation*}
    q_{n,j} = \myvec{q}_{rad,j} \cdot \myvec{\hat{n}}_j = 
    \int_{4\pi} I_j \myvec{\hat{n}}_j\cdot\myvec{\hat{s}} d\Omega  \approx \sum_{k=1}^{N_s} I_{j,k} (\myvec{\hat{n}}_j \cdot \myvec{\hat{s}}_k) \Delta \Omega_k.
\end{equation*}

To provide a more accurate estimate of the true integral for $\dot{Q}_{out, f}$, we employ the 2D trapezoidal rule, given by the formula $\dot{Q}_{out, f} \approx \frac{\Delta S}{4} \sum_{i=1}^{M} \sum_{j=1}^{N} w_{i,j} \, q_{n, i, j}$. The formula's double summation iterates over the $M \times N$ grid on the face, where $q_{n, i, j}$ is the local normal heat flux and $w_{i,j}$ is the trapezoidal weight determined by a point's location: 1 for corners, 2 for edges, and 4 for all interior points.

As mentioned in Section~\ref{sec:datasets}, we have $N_s = 16$ solid angles. For the 3D McCaffrey case, the step sizes for the azimuthal angle $\phi$ and polar angle $\theta$ are $\Delta \phi = \frac{\pi}{2 n_{\phi}}$ and $\Delta \theta = \frac{\pi}{ n_{\theta}}$. Then the angles are ($\theta_n = \frac{(2n+1)\Delta \theta}{2}$, $\phi_m = \frac{(2m+1)\Delta \phi}{2}$), where $m = 0,1,\ldots,4n_{\phi}-1$, $n=0,1,\ldots,n_{\theta}-1$, and $4n_{\phi}n_{\theta} = N_s$. The unit direction vectors are given by $\myvec{\hat{s}}{n,m} = (\sin\theta_n \sin\phi_m, \sin\theta_n \cos\phi_m, \cos\theta_n)$, and the discrete solid angle is $\Delta\Omega{n,m} = \sin(\theta_n) \Delta\theta \Delta\phi$. The dot product $\myvec{\hat{n}}f \cdot \myvec{\hat{s}}{n,m}$ depends on the orientation of the boundary face. By substituting the discrete approximation of $q_{n,j}$, we have $ \dot{Q}{out,f} \approx \sum{j=1}^{N_f} \left( \sum_{n=0}^{n_{\theta}-1} \sum_{m=0}^{4n_{\phi}-1} I_{j,n,m} (\mathbf{\hat{n}}f \cdot \mathbf{\hat{s}}{n,m}) \Delta\Omega_{n,m} \right) \Delta S_j$.

\paragraph{Radiation fraction}
The radiation fraction is the ratio of the radiative loss to the chemical reaction rate. It is an integral quantity that measures the percentage of energy released from the combustion process lost via radiation:
\begin{equation*}
    \chi_R = \frac{\dot{Q}_{ems} - \dot{Q}_{abs}}{\dot{Q}_{chem}},
\end{equation*}
where $\dot{Q}_{ems} = \int_V 4\pi \kappa I_b dV$, and $\dot{Q}_{abs} = \int_V \kappa \left(\int_{4\pi} I d\Omega \right) dV$.

\paragraph{Energy conservation}
The divergence of the radiative heat flux is
\begin{eqnarray*}
\nabla\cdot \myvec{q}_{rad} &=& \nabla\cdot\int{4\pi} I \myvec{\hat{s}} d\Omega = \int_{4\pi} \myvec{\hat{s}} \cdot \nabla I d\Omega \\
&=& \int_{4\pi} \left( \kappa I_{b} - \kappa I \right) d\Omega \\
&=& \underbrace{4\pi \kappa I_b}_{\text{Emission}} - \underbrace{\kappa \int{4\pi} I d\Omega}_{\text{Absorption}} = \kappa (4\pi I_b - G),
\end{eqnarray*}
where $G = \int{4\pi} I d\Omega$ is the incident radiation. Based on the divergence theorem, the energy conservation should be 
$$\underbrace{\oint_{S} (\myvec{q}_{rad} \cdot \myvec{\hat{n}}) dS}_{\text{Net radiative heat loss}} = \underbrace{\int_{V} (\nabla \cdot \myvec{q}_{rad})dV}_{\text{Net radiative power source}}.$$

\section{Fourier-MIONet architectures}
\label{appx:hyperparameters}

The network architectures of nested Fourier-MIONets for all levels in Section~\ref{sec:nested_fourier-mionet} are summarized in Table~\ref{tab:arch_nested}.  The trunk net activation is Swish (SiLU). 

\begin{table}[htbp]
\centering
\footnotesize
\caption{\textbf{Architecture of the nested Fourier-MIONet across all four levels.} $C$ denotes the batch size, $W$ the network width, and $N$ the number of Fourier layers.}
\label{tab:arch_nested}
\begin{tabular}{llll}
\toprule
Level &  & Operation & Output Shape \\
\midrule
\multirow{9}{*}{Level 4} & Branch net 1 & Linear(4, W), Permute & ($C$, $W$, 30, 30, 30) \\
 & Branch net 2 & Linear(4, W), Permute & ($C$, $W$, 30, 30, 30) \\
 & Trunk net & FNN [2, $W$, $W$, $W$] & (16, $W$) \\ 
 & Branch merger & Point-wise addition & ($C$, $W$, 30, 30, 30) \\
 & Branch-trunk merger & Point-wise multiplication & ($C$, 16, $W$, 30, 30, 30) \\ 
 & \multirow{3}{*}{Merge net} & $N$ Fourier layers & ($C \times 16$, $W$, 30, 30, 30) \\
 &  & Linear($W$, $W \times 4$), ReLU & ($C \times 16$, 30, 30, 30, $W \times 4$) \\
 &  & Linear($W \times 4$, 1) & ($C \times 16$, 30, 30, 30, 1) \\ 
 & Reshape & Reshape, Output transform & ($C, 16 \times 30 \times 30 \times 30$) \\
\midrule

\multirow{9}{*}{Level 3} & Branch net 1 & CNN & ($C$, $W$, 24, 24, 40) \\
 & Branch net 2 & Linear(5, W), Permute & ($C$, $W$, 24, 24, 40) \\
 & Trunk net & FNN [2, $W$, $W$, $W$] & (16, $W$) \\
 & Branch merger & Point-wise multiplication & ($C$, $W$, 24, 24, 40) \\
 & Branch-trunk merger & Point-wise multiplication & ($C$, 16, $W$, 24, 24, 40) \\
 & \multirow{3}{*}{Merge net} & $N$ Fourier layers & ($C \times 16$, $W$, 24, 24, 40) \\
 &  & Linear($W$, $W \times 4$), ReLU & ($C \times 16$, 24, 24, 40, $W \times 4$) \\
 &  & Linear($W \times 4$, 1) & ($C \times 16$, 24, 24, 40, 1) \\
 & Reshape & Reshape, Output transform & ($C, 16 \times 24 \times 24 \times 40$) \\
\midrule

\multirow{9}{*}{Level 2} & Branch net 1 & CNN & ($C$, $W$, 40, 40, 40) \\
 & Branch net 2 & Linear(5, W), Permute & ($C$, $W$, 40, 40, 40) \\
 & Trunk net & FNN [2, $W$, $W$, $W$] & (16, $W$) \\
 & Branch merger & Point-wise multiplication & ($C$, $W$, 40, 40, 40) \\
 & Branch-trunk merger & Point-wise multiplication & ($C$, 16, $W$, 40, 40, 40) \\
 & \multirow{3}{*}{Merge net} & $N$ Fourier layers & ($C \times 16$, $W$, 40, 40, 40) \\
 &  & Linear($W$, $W \times 4$), ReLU & ($C \times 16$, 40, 40, 40, $W \times 4$) \\
 &  & Linear($W \times 4$, 1) & ($C \times 16$, 40, 40, 40, 1) \\
 & Reshape & Reshape, Output transform & ($C, 16 \times 40 \times 40 \times 40$) \\
\midrule

\multirow{9}{*}{Level 1} & Branch net 1 & CNN & ($C$, $W$, 48, 48, 80) \\
 & Branch net 2 & Linear(5, W), Permute & ($C$, $W$, 48, 48, 80) \\
 & Trunk net & FNN [2, $W$, $W$, $W$] & (16, $W$) \\
 & Branch merger & Point-wise multiplication & ($C$, $W$, 48, 48, 80) \\
 & Branch-trunk merger & Point-wise multiplication & ($C$, 16, $W$, 48, 48, 80) \\
 & \multirow{3}{*}{Merge net} & $N$ Fourier layers & ($C \times 16$, $W$, 48, 48, 80) \\
 &  & Linear($W$, $W \times 4$), ReLU & ($C \times 16$, 48, 48, 80, $W \times 4$) \\
 &  & Linear($W \times 4$, 1) & ($C \times 16$, 48, 48, 80, 1) \\
 & Reshape & Reshape, Output transform & ($C, 16 \times 48 \times 48 \times 80$) \\
\bottomrule
\end{tabular}
\end{table}

\section{Additional results on 3D McCaffrey fires}
\label{appx:fixed_HRR}

In Section~\ref{sec:fixed_HRR}, we train the nested Fourier-MIONet on five McCaffrey fires with fixed HRRs of 14, 22, 33, 45, and 58 kW using four model sizes: Tiny, Small, Medium, and Large, which differ in network width, the number of retained Fourier modes, and the depth of Fourier layers (Table~\ref{tab:MIONet_config}).
We show the additional global testing results of fire with HRR 22, 33, and $45~\mathrm{kW}$ in Table~\ref{tab:McCaffrey_22_33_45_bottom}. The per-level $L^2$ relative errors and SSIM values for radiative intensity $I$ and incident radiation $G$ during training are reported in Table~\ref{tab:McCaffrey_train_IG}. For both the radiative intensity $I$ and the incident radiation $G$, the global relative errors decrease as the model size increases from Tiny to Large, indicating improved global accuracy with increasing capacity. At a fixed model size, higher-HRR cases tend to be more challenging and yield larger errors. 

\begin{table*}[htbp]
  \centering
  \caption{\textbf{Model and per-level configuration of the nested Fourier-MIONet used for the 3D McCaffrey fires.} Four model sizes are considered: Tiny, Small, Medium, and Large, which differ in the network width (``Width'') of the branch, trunk, and Fourier decoder networks, the number of retained Fourier modes (``Modes'') in the three spatial dimensions, and the number of Fourier layers. The architecture and hyperparameters are shared across all sizes of fire (14--58~kW).}
  \label{tab:MIONet_config}
  \begin{minipage}{\textwidth}
  \centering
  \footnotesize
  \setlength{\tabcolsep}{6.0pt}
  \begin{tabular}{c c c c c c c}
    \toprule
    \multirow{2}{*}{\textbf{Model}} &
    \multirow{2}{*}{\textbf{Level}} &
    \multirow{2}{*}{\textbf{Width}} &
    \multicolumn{2}{c}{\textbf{Modes}} &
    \multirow{2}{*}{\#\,\textbf{Fourier layers}} &
    \multirow{2}{*}{\#\,\textbf{Params}} \\
    \cmidrule(lr){4-5}
      & & & 1\&2 & 3 & & \\
    \midrule
% ---------- Model Tiny ----------
    \multirow{4}{*}{Tiny}
      & 4 & 16 &  4 & 4 & 1 &   67713 \\
      & 3 & 16 &  4 & 4 & 1 &  176326 \\
      & 2 & 16 &  4 & 4 & 1 &  127211 \\
      & 1 & 16 &  4 & 4 & 1 &  131182 \\
    \midrule
% ---------- Model Small ----------
    \multirow{4}{*}{Small}
      & 4 & 24 &  8 & 4 & 2 & 1184857 \\
      & 3 & 24 &  8 & 4 & 2 & 1293598 \\
      & 2 & 24 &  8 & 4 & 2 & 1244483 \\
      & 1 & 24 &  8 & 4 & 2 & 1248454 \\
    \midrule
% ---------- Model Medium ----------
    \multirow{4}{*}{Medium}
      & 4 & 32 &  8 & 8 & 2 & 4203297 \\
      & 3 & 32 &  8 & 8 & 2 & 4312166 \\
      & 2 & 32 &  8 & 8 & 2 & 4263051 \\
      & 1 & 32 &  8 & 8 & 2 & 4267022 \\
    \midrule
% ---------- Model Large ----------
    \multirow{4}{*}{Large}
      & 4 & 48 & 12 & 8 & 3 & 31872481 \\
      & 3 & 48 & 12 & 8 & 3 & 31981606 \\
      & 2 & 48 & 12 & 8 & 3 & 31932491 \\
      & 1 & 48 & 12 & 8 & 3 & 31936462 \\
    \bottomrule
  \end{tabular}
  \end{minipage}
\end{table*}

\begin{table*}[htbp]
  \centering
  \caption{\textbf{Per-level and global testing performance of the nested Fourier-MIONet for the 3D McCaffrey 22, 33, and 45~kW fires.} We report the individual per-level errors and the global errors obtained during nested inference. The reference radiative fractions $\chi_R$ are 19.40\%, 20.91\%, and 21.76\% for the 22, 33, and 45 kW fires, respectively.}
  \label{tab:McCaffrey_22_33_45_bottom}
  \footnotesize
  \setlength{\tabcolsep}{6.0pt}
  \begin{tabular}{c c cccc c c c c c c}
    \toprule
    \multirow{2}{*}{\textbf{Fire}} &
    \multirow{2}{*}{\textbf{Model}} &
    \multicolumn{4}{c}{\textbf{Individual error $\varepsilon^{\text{level}}_I$}} &
    \multirow{2}{*}{$\varepsilon^{\text{global}}_I$} &
    \multirow{2}{*}{$\varepsilon^{\text{global}}_G$} &
    \multirow{2}{*}{$\varepsilon_{\dot{Q}_{out}}$} &
    \multirow{2}{*}{$\varepsilon_{\chi_R}$} &
    \multirow{2}{*}{$\mathcal{R}_E^{pred}$} &
    \multirow{2}{*}{$\mathcal{R}_E^{ref}$} \\
    \cmidrule(lr){3-6}
      & & Level 4 & Level 3 & Level 2 & Level 1 & & & & & & \\
    \midrule

    \multirow{4}{*}{22 kW}
      & Tiny  & 5.52\% & 10.74\% & 7.26\% & 6.85\% & 2.99\% & 1.35\% & 0.19\% & 0.19\% & 0.38$~\mathrm{kW}$ & \multirow{4}{*}{0.45$~\mathrm{kW}$} \\
      & Small & 4.26\% &  8.28\% & 4.95\% & 4.86\% & 2.03\% & 1.14\% & 0.15\% & 0.15\% & 0.42$~\mathrm{kW}$ & \\
      & Medium  & 3.69\% &  6.15\% & 3.91\% & 2.81\% & 1.64\% & 0.98\% & 0.13\% & 0.13\% & 0.44$~\mathrm{kW}$ & \\
      & Large & 2.43\% &  2.94\% & 2.25\% & 1.43\% & 1.04\% & 0.60\% & 0.03\% & 0.04\% & 0.44$~\mathrm{kW}$ & \\

    \midrule

    \multirow{4}{*}{33 kW}
      & Tiny  & 7.93\% & 13.93\% & 8.78\% & 8.40\% & 4.28\% & 2.02\% & 0.16\% & 0.16\% & 0.20$~\mathrm{kW}$ & \multirow{4}{*}{0.37$~\mathrm{kW}$} \\
      & Small & 6.14\% &  9.66\% & 5.31\% & 5.12\% & 2.89\% & 1.60\% & 0.13\% & 0.13\% & 0.36$~\mathrm{kW}$ & \\
      & Medium  & 4.92\% &  6.51\% & 4.11\% & 3.05\% & 2.26\% & 1.31\% & 0.21\% & 0.21\% & 0.40$~\mathrm{kW}$ & \\
      & Large & 3.77\% &  4.67\% & 3.05\% & 1.98\% & 1.53\% & 0.76\% & 0.04\% & 0.04\% & 0.38$~\mathrm{kW}$ & \\

    \midrule

    \multirow{4}{*}{45 kW}
      & Tiny   & 9.76\% & 15.19\% & 10.16\% & 9.44\% & 5.42\% & 2.66\% & 0.13\% & 0.13\% & 0.15$~\mathrm{kW}$ & \multirow{4}{*}{0.15$~\mathrm{kW}$} \\
      & Small  & 7.04\% & 10.16\% &  5.65\% & 5.60\% & 3.51\% & 1.81\% & 0.12\% & 0.12\% & 0.13$~\mathrm{kW}$ & \\
      & Medium & 4.89\% &  5.78\% &  3.63\% & 2.59\% & 2.48\% & 1.39\% & 0.14\% & 0.13\% & 0.18$~\mathrm{kW}$ & \\
      & Large  & 2.67\% &  2.73\% &  2.17\% & 1.58\% & 1.15\% & 0.44\% & 0.00\% & 0.00\% & 0.14$~\mathrm{kW}$ & \\

    \bottomrule
  \end{tabular}
\end{table*}

\begin{table*}[htbp]
  \centering
  \caption{\textbf{Per-level testing $L^2$ relative errors and SSIM values for radiative intensity $I$ and incident radiation $G$ during training on McCaffrey fires of 14, 22, 33, 45, and 58 kW.} The upper block reports results for $I$, and the lower block reports results for $G$. Model configurations are provided in Table~\ref{tab:MIONet_config}.}
  \label{tab:McCaffrey_train_IG}
  \footnotesize
  \setlength{\tabcolsep}{6.0pt}
  \resizebox{0.9\textwidth}{!}{
  \begin{tabular}{c c cc cc cc cc cc}
    \toprule
    \multicolumn{12}{c}{Radiative intensity $I$} \\
    \midrule
    \multirow{2}{*}{\textbf{Model}} &
    \multirow{2}{*}{\textbf{Level}} &
    \multicolumn{2}{c}{\textbf{14 kW}} &
    \multicolumn{2}{c}{\textbf{22 kW}} &
    \multicolumn{2}{c}{\textbf{33 kW}} &
    \multicolumn{2}{c}{\textbf{45 kW}} &
    \multicolumn{2}{c}{\textbf{58 kW}} \\
    \cmidrule(lr){3-4}\cmidrule(lr){5-6}\cmidrule(lr){7-8}\cmidrule(lr){9-10}\cmidrule(lr){11-12}
      & & $\varepsilon_I$ & SSIM$_I$ & $\varepsilon_I$ & SSIM$_I$ & $\varepsilon_I$ & SSIM$_I$ & $\varepsilon_I$ & SSIM$_I$ & $\varepsilon_I$ & SSIM$_I$ \\
    \midrule
    \multirow{4}{*}{Tiny}
      & 4 & 3.69\% & 0.996 & 5.52\% & 0.994 & 8.76\% & 0.921 & 9.75\% & 0.928 & 10.50\% & 0.986 \\
      & 3 & 5.08\% & 0.995 & 6.28\% & 0.994 & 9.01\% & 0.898 & 9.54\% & 0.908 & 10.32\% & 0.983 \\
      & 2 & 5.24\% & 0.993 & 5.24\% & 0.993 & 8.41\% & 0.816 & 9.88\% & 0.795 & 9.48\% & 0.972 \\
      & 1 & 5.52\% & 0.986 & 6.16\% & 0.981 & 7.53\% & 0.966 & 8.60\% & 0.953 & 9.05\% & 0.949 \\
    \midrule
    \multirow{4}{*}{Small}
      & 4 & 3.00\% & 0.998 & 4.26\% & 0.997 & 5.90\% & 0.973 & 7.07\% & 0.957 & 7.68\% & 0.993 \\
      & 3 & 3.21\% & 0.998 & 4.53\% & 0.997 & 5.54\% & 0.957 & 6.66\% & 0.954 & 6.82\% & 0.994 \\
      & 2 & 3.08\% & 0.998 & 3.53\% & 0.997 & 4.34\% & 0.893 & 5.13\% & 0.890 & 4.73\% & 0.993 \\
      & 1 & 2.71\% & 0.996 & 3.54\% & 0.993 & 4.25\% & 0.988 & 5.03\% & 0.983 & 5.11\% & 0.984 \\
    \midrule
    \multirow{4}{*}{Medium}
      & 4 & 2.50\% & 0.999 & 3.69\% & 0.997 & 3.95\% & 0.982 & 4.92\% & 0.973 & 5.39\% & 0.995 \\
      & 3 & 2.11\% & 0.999 & 2.97\% & 0.999 & 3.45\% & 0.971 & 4.25\% & 0.968 & 4.02\% & 0.997 \\
      & 2 & 2.42\% & 0.999 & 2.72\% & 0.998 & 2.91\% & 0.930 & 3.17\% & 0.931 & 3.15\% & 0.997 \\
      & 1 & 1.72\% & 0.998 & 1.82\% & 0.997 & 2.30\% & 0.995 & 2.25\% & 0.995 & 2.39\% & 0.994 \\
    \midrule
    \multirow{4}{*}{Large}
      & 4 & 1.66\% & 0.999 & 2.43\% & 0.999 & 2.19\% & 0.992 & 2.71\% & 0.992 & 2.63\% & 0.999 \\
      & 3 & 1.54\% & 1.000 & 1.82\% & 0.999 & 2.00\% & 0.989 & 2.39\% & 0.989 & 2.28\% & 0.999 \\
      & 2 & 1.65\% & 0.999 & 1.90\% & 0.999 & 1.98\% & 0.944 & 2.11\% & 0.968 & 2.12\% & 0.999 \\
      & 1 & 1.67\% & 0.997 & 1.32\% & 0.998 & 1.32\% & 0.998 & 1.52\% & 0.998 & 2.02\% & 0.995 \\
    \midrule
    \multicolumn{12}{c}{Incident radiation $G$} \\
    \midrule
    \multirow{2}{*}{\textbf{Model}} &
    \multirow{2}{*}{\textbf{Level}} &
    \multicolumn{2}{c}{\textbf{14 kW}} &
    \multicolumn{2}{c}{\textbf{22 kW}} &
    \multicolumn{2}{c}{\textbf{33 kW}} &
    \multicolumn{2}{c}{\textbf{45 kW}} &
    \multicolumn{2}{c}{\textbf{58 kW}} \\
    \cmidrule(lr){3-4}\cmidrule(lr){5-6}\cmidrule(lr){7-8}\cmidrule(lr){9-10}\cmidrule(lr){11-12}
      & & $\varepsilon_G$ & SSIM$_G$ & $\varepsilon_G$ & SSIM$_G$ & $\varepsilon_G$ & SSIM$_G$ & $\varepsilon_G$ & SSIM$_G$ & $\varepsilon_G$ & SSIM$_G$ \\
    \midrule
    \multirow{4}{*}{Tiny}
      & 4 & 1.37\% & 0.997 & 2.21\% & 0.992 & 3.51\% & 0.972 & 4.11\% & 0.970 & 4.32\% & 0.976 \\
      & 3 & 1.69\% & 0.998 & 2.02\% & 0.997 & 3.05\% & 0.949 & 3.42\% & 0.960 & 3.58\% & 0.991 \\
      & 2 & 1.96\% & 0.998 & 1.96\% & 0.998 & 3.28\% & 0.988 & 3.95\% & 0.985 & 3.76\% & 0.991 \\
      & 1 & 1.99\% & 0.995 & 2.28\% & 0.994 & 2.94\% & 0.990 & 3.44\% & 0.988 & 3.54\% & 0.988 \\
    \midrule
    \multirow{4}{*}{Small}
      & 4 & 1.19\% & 0.997 & 1.78\% & 0.994 & 2.53\% & 0.985 & 2.90\% & 0.978 & 3.07\% & 0.988 \\
      & 3 & 1.09\% & 0.999 & 1.50\% & 0.999 & 1.97\% & 0.976 & 2.43\% & 0.975 & 2.55\% & 0.996 \\
      & 2 & 1.25\% & 0.999 & 1.46\% & 0.999 & 1.82\% & 0.996 & 2.16\% & 0.996 & 2.00\% & 0.998 \\
      & 1 & 1.02\% & 0.999 & 1.38\% & 0.998 & 1.70\% & 0.996 & 2.04\% & 0.996 & 2.08\% & 0.996 \\
    \midrule
    \multirow{4}{*}{Medium}
      & 4 & 0.97\% & 0.998 & 1.54\% & 0.995 & 1.82\% & 0.989 & 2.09\% & 0.983 & 2.24\% & 0.991 \\
      & 3 & 0.75\% & 1.000 & 1.06\% & 0.999 & 1.25\% & 0.987 & 1.55\% & 0.986 & 1.54\% & 0.998 \\
      & 2 & 0.99\% & 1.000 & 1.14\% & 0.999 & 1.24\% & 0.998 & 1.34\% & 0.998 & 1.36\% & 0.999 \\
      & 1 & 0.67\% & 0.999 & 0.71\% & 0.999 & 0.93\% & 0.999 & 0.91\% & 0.999 & 0.99\% & 0.999 \\
    \midrule
    \multirow{4}{*}{Large}
      & 4 & 0.71\% & 0.998 & 0.99\% & 1.000 & 0.80\% & 0.995 & 0.93\% & 0.995 & 0.94\% & 0.999 \\
      & 3 & 0.55\% & 1.000 & 0.69\% & 1.000 & 0.77\% & 0.995 & 0.94\% & 0.994 & 0.94\% & 0.999 \\
      & 2 & 0.67\% & 1.000 & 0.84\% & 1.000 & 0.74\% & 0.998 & 0.96\% & 0.999 & 0.98\% & 0.999 \\
      & 1 & 0.69\% & 0.999 & 0.55\% & 1.000 & 0.54\% & 0.999 & 0.64\% & 1.000 & 0.90\% & 0.999 \\
    \bottomrule
  \end{tabular}}
\end{table*}

In Section~\ref{sec:variable_HRR}, we train the nested Fourier-MIONet using four model sizes (Tiny, Small, Medium, and Large) on McCaffrey fires with variable HRRs, and we report the individual per-level errors and the global errors obtained during nested inference for each HRR. Here, the per-level $L^2$ relative errors and SSIM values for radiative intensity $I$ and incident radiation $G$ during training are detailed in Table~\ref{tab:McCaffrey_all_training}.

\begin{table*}[htbp]
  \centering
  \caption{\textbf{Per-level testing $L^2$ relative errors and SSIM values for radiative intensity $I$ and incident radiation $G$ when training the entire nested Fourier-MIONet on variable HRRs.}}
  \label{tab:McCaffrey_all_training}
  \begin{minipage}{\textwidth}
  \centering
  \footnotesize
  \begin{tabular}{c c c c c c}
    \toprule
    \multirow{2}{*}{\textbf{Model}} &
    \multirow{2}{*}{\textbf{Level}} &
    \multicolumn{2}{c}{\textbf{$I$}} &
    \multicolumn{2}{c}{\textbf{$G$}} \\ 
    \cmidrule(lr){3-4} \cmidrule(l){5-6}
     & & $\varepsilon_I$ & SSIM$_I$ & $\varepsilon_G$ & SSIM$_G$ \\
    \midrule

% ---------- Model Tiny ----------
    \multirow{4}{*}{Tiny}
      & 4 & 8.16\% & 0.989 & 3.54\% & 0.974\\
      & 3 & 9.59\% & 0.986 & 3.95\% & 0.988\\
      & 2 & 8.94\% & 0.970 & 4.01\% & 0.986\\
      & 1 & 8.12\% & 0.925 & 4.33\% & 0.962 \\
    \midrule

% ---------- Model Small ----------
    \multirow{4}{*}{Small}
      & 4 & 7.66\% & 0.993 & 3.17\% & 0.985\\
      & 3 & 7.55\% & 0.987 & 3.77\% & 0.977\\
      & 2 & 4.37\% & 0.994 & 2.01\% & 0.997\\
      & 1 & 4.55\% & 0.981 & 2.47\% & 0.984 \\
    \midrule

% ---------- Model Medium ----------
    \multirow{4}{*}{Medium}
      & 4 & 6.80\% & 0.994 & 2.91\% & 0.985 \\
      & 3 & 4.63\% & 0.994 & 2.09\% & 0.991 \\
      & 2 & 3.25\% & 0.994 & 1.84\% & 0.994 \\
      & 1 & 3.85\% & 0.968 & 3.25\% & 0.964 \\
    \midrule

% ---------- Model Large ----------
    \multirow{4}{*}{Large}
      & 4 & 6.72\% & 0.994 & 3.00\% & 0.986 \\
      & 3 & 5.61\% & 0.996 & 2.52\% & 0.995 \\
      & 2 & 2.68\% & 0.997 & 1.60\% & 0.997 \\
      & 1 & 3.65\% & 0.984 & 3.07\% & 0.983 \\
    \bottomrule
  \end{tabular}
  \end{minipage}
\end{table*}